\documentclass[aps,prd,superscriptaddress,showkeys,showpacs,twocolumn,a4paper]{revtex4-1}
\usepackage{ulem}
\usepackage{color}
\usepackage{graphicx}
\usepackage{appendix}
\usepackage{epsfig}
\usepackage[colorlinks=true,linkcolor=blue]{hyperref}
\usepackage{epstopdf}
\usepackage{amsmath}
\usepackage{subfig}
\usepackage{comment}
\usepackage[dvipsnames]{xcolor}
\newcommand{\be}{\begin{equation}}
\newcommand{\ee}{\end{equation}}
\newcommand{\ba}{\begin{eqnarray}}
\newcommand{\ea}{\end{eqnarray}}

\begin{document}
\title{Response of a weakly magnetized hot QCD medium to inhomogeneous electric field}

\author{Gowthama K K}
\email{k$_$gowthama@iitgn.ac.in}
\affiliation{Indian Institute of Technology Gandhinagar, Gandhinagar-382355, Gujarat, India}

\author{Manu Kurian}
\email{manu.kurian@iitgn.ac.in}
\affiliation{Indian Institute of Technology Gandhinagar, Gandhinagar-382355, Gujarat, India}

\author{Vinod Chandra}
\email{vchandra@iitgn.ac.in}
\affiliation{Indian Institute of Technology Gandhinagar, Gandhinagar-382355, Gujarat, India}

\begin{abstract}
The electric charge transport in a weakly magnetized hot QCD medium has been investigated in the presence of an external inhomogeneous electric field. The current densities (electric and Hall) induced by the inhomogeneous electric field have been estimated in the regime where space-time inhomogeneity of the field is small so that the collisional effect in the medium cannot be neglected. The collisional aspect of the medium has been captured by employing both relaxation time approximation and Bhatnagar-Gross-Krook collision kernel in the relativistic Boltzmann equation. The magnetic field, momentum anisotropy, and quark chemical potential dependences of the electric current and Hall current densities have been explored, and the impacts on the respective conductivities have been studied. The inhomogeneities of the field are seen to have sizable effects on the electromagnetic responses of the collisional medium.

\end{abstract}
\maketitle

 \section{Introduction}
The recent Large Hadron Collider (LHC) and Relativistic Heavy-Ion Collider (RHIC)  observations on the directed flow for charged hadrons and $D/\bar{D}^0$ mesons provide insights into the existence of a strong electromagnetic field in the early stages of heavy-ion collision~\cite{Acharya:2019ijj,Adam:2019wnk}. The strength of the created magnetic field is estimated to be the order of $(1-15)m_\pi^2$ in the initial stages of collision~\cite{Skokov:2009qp,Zhong:2014cda,Voronyuk:2011jd,Deng:2012pc}. The study of deconfined nuclear matter- quark gluon plasma (QGP) in electromagnetic fields and the associated phenomenological aspects has gained huge momentum over the last decade~\cite{Fukushima:2008xe,Kharzeev:2015znc,Gusynin:1995nb,Sadofyev:2010pr,Bali:2011qj,Kurian:2020kct,Karmakar:2018aig,Ghosh:2019kmf,Greif:2017irh,Singh:2017nfa,Feng:2017giy,Rath:2020idp}. However, a proper theory/model describing the evolution of the electromagnetic fields in the QGP is not yet  formulated completely. Various studies have revealed that the evolution of the fields may depend on the hot QCD medium properties whereas the fields decay rapidly in the vacuum~\cite{Tuchin:2013apa,Tuchin:2015oka,McLerran:2013hla}.

To describe the impact of the generated fields on the system of quarks/antiquarks and gluons, electromagnetic responses in the medium play quite an important role~\cite{Zakharov:2014dia,Gursoy:2014aka,Jiang:2020lgw}. The electrical conductivity quantifies the electric current being induced in response to the fields and also controls the late time behaviour of the fields in the electromagnetically charged medium. There have been various studies to understand the electrical conductivity of the QCD medium from transport theory~\cite{Greif:2014oia,Cassing:2013iz,Puglisi:2014sha,Mitra:2016zdw,Thakur:2017hfc}, lattice QCD computations~\cite{Astrakhantsev:2019zkr,Amato:2013naa,Francis:2011bt}, Kubo formalism~\cite{Kubo:1957mj}, and holographic methods~\cite{Finazzo:2015xwa,Jain:2010ip}. Further, the electrical conductivity has been extracted from the soft photon spectrum and charge-dependent flow coefficients from heavy-ion collisions in Refs.~\cite{Yin:2013kya,Hirono:2012rt}. In a weakly magnetized medium, the temperature is the dominant energy scale of the system in comparison to the strength of the magnetic field. On the other hand, the system follows $1+1-$dimensional Landau level kinematics in the presence of the strong magnetic field. The longitudinal electrical conductivity in a strongly magnetized medium has been explored recently~\cite{Hattori:2016cnt,Hattori:2016lqx,Fukushima:2017lvb,Kurian:2017yxj,Ghosh:2019ubc}. In Refs.~\cite{Feng:2017tsh,Das:2019ppb,Das:2019wjg,Thakur:2019bnf,Chatterjee:2019nld,Dey:2019axu,Kurian:2020qjr,Kalikotay:2020snc}, the dominant components of current density in various directions and the associated transport coefficients, electrical and Hall conductivities, have been studied in a weakly magnetized medium. Those investigations assume the electrostatic fields in the medium as constant. Several studies have revealed the possibility of the generation of inhomogeneous electromagnetic fields in the heavy-ion collision experiments~\cite{Deng:2012pc,Tuchin:2013apa,Hongo:2013cqa}. Therefore, it is an interesting aspect to study the responses of the QGP medium in the presence of inhomogeneous fields.

The collisional aspects are incorporated in the present analysis by choosing the regime of focus where space-time inhomogeneity of the fields is small so that the collision effects could be significant. The collisional effects are embedded in the estimation of electromagnetic responses via relaxation-time approximation (RTA)~\cite{anderson1974relativistic} and Bhatnagar-Gross-Krook (BGK) collision kernel~\cite{Bhatnagar:1954zz} while solving the relativistic Boltzmann equation. The difference between the conventional RTA and BGK collisional kernel is that in the later approach the particle number is conserved instantaneously.

In the current study, we have estimated the electric and Hall current in the presence of inhomogeneous fields. The near-equilibrium distribution function due to the fields has been obtained by solving the Boltzmann equation with RTA and BGK collision kernels. The dependence of the quark chemical potential and strength of the magnetic field on the Hall current has also been explored in the analysis. Besides, the medium is also analyzed in the presence of weak anisotropy to get insights on the impact of the anisotropy on the charge transport in the hot QCD medium. We have illustrated the effects of inhomogeneity of the fields and momentum anisotropy of the medium through the temperature behaviour of current densities in a weakly magnetized collisional medium.

The present manuscript is organized as follows. In Section II, the formalism for electric charge transport in the weakly magnetized medium is described within the RTA and BGK collision kernels in the presence of an inhomogeneous electric field along with the analysis on weak momentum anisotropy. Section III is devoted to discussions on the effects of inhomogeneity of the electric field and the collisional aspects of the medium to the electromagnetic responses of a weakly magnetized QGP. Finally, we conclude the analysis with an outlook in Section IV.

{\bf Notations and conventions:} The subscript $k$ denotes the particle species, $k = (q, \bar{q})$, where $ q$ and $\bar{q}$ represent quarks and antiquarks, respectively. The quantity $q_{f_k}$ is the charge of the particle of flavor $f$, $N_f=3$ is the number of flavours, $N_c$ is the number of colors, and $g_k=2N_c\sum_f$ implies the degeneracy factor of the $k-$th species. The fluid velocity $u^{\mu} = (1, 0, 0, 0)$ is normalized to unity in the rest frame ($u^{\mu} u_{\mu} =1$). The quantities $B=\mid{\bf B}\mid$ and $E=\mid{\bf E}\mid$ represent the magnitude of magnetic and electric fields.

\section{Formalism: Electric and Hall current in a weakly magnetized QCD medium}
The magnetic field breaks the rotational symmetry and induces anisotropy to the hot QCD medium leading to anisotropic transport processes in the medium. Studies have shown that the momentum transport in the presence of a weak magnetic field gives rise to five components of shear tensor and two components of bulk viscous pressure in the medium~\cite{Dash:2020vxk,Ghosh:2020wqx,Panda:2020zhr,Huang:2011dc,Denicol:2019iyh}. To quantify the effects of the magnetic field in the electric charge transport, one needs to study the current density in the magnetized medium. The first step towards the estimation of current density is to obtain the non-equilibrium part of the distribution function from the relativistic Boltzmann equation. In the magnetized medium, the Boltzmann equation can be defined as,
\begin{align}\label{1}
{p}^{\mu}\,\partial_{\mu}f_k(x,{p})+q_{f_k}F^{\mu\nu}{p}_{  \nu}\partial^{(p)}_{\mu} f_k=C(f_k),
\end{align}
where $F^{\mu\nu}$ is the electromagnetic field strength tensor, $C(f_k)$ is the collision kernel near equilibrium and $f_k=f^0_k+\delta f_k$ is the distribution function for quarks and antiquarks at finite quark chemical potential $\mu$ with equilibrium distribution as
\begin{equation}
f^0_k=\frac{1}{1+\exp{\big(\beta( \epsilon \mp \mu)\big)}}.
\end{equation}
The non-equilibrium part of the distribution
function can be obtained by solving the Boltzmann equation by choosing the ansatz as follows,
\begin{equation}\label{2}
\delta f_k=({\bf{p}}.{\bf \Xi} ) \frac{\partial f^0_k}{\partial \epsilon}, 
\end{equation}
where the vector ${\bf{\Xi}}$ is related to the electric and the magnetic field in the medium. The present analysis incorporates the effects of the spacetime inhomogeneity of the fields in the electromagnetic responses of the medium. These inhomogeneities generate additional 
components to the current density and can be described in terms of spacetime derivatives of the fields. 
Hence, the vector ${\bf{\Xi}}$ takes the following form considering the terms up to first order derivative of the fields,
\begin{align}\label{3}
\mathbf{\Xi} =& \alpha_1\textbf{E}+ \alpha_2\dot{\textbf{E}}+ \alpha_3(\textbf{E}\times \textbf{B})+ \alpha_4(\dot{\textbf{E}}\times \textbf{B})+ \alpha_5(\textbf{E}\times \dot{\textbf{B}})\nonumber\\&+\alpha_6 ({\nabla} \times \textbf{E}) +\alpha_7 \textbf{B}+\alpha_8 \dot{\textbf{B}}+\alpha_9 ({\nabla} \times \textbf{B}),
\end{align} 
where $\alpha_{i}$ ($i=(1, 2,.., 9)$) are the unknown functions that need to be calculated from the microscopic description. It is important to emphasize that the parity of the current operator is different from that of ${\bf B}$, $\dot{\bf B}$ and ${\bf \nabla}\times{\bf E}$. Hence, the components of the current density with these vector quantities cannot exist as they violate parity considerations $i.e.$, $\alpha_{i}=0$ for $i=(6,7,8)$. However, these components will have finite contributions to the current density for a system with a non-zero chiral chemical potential $\mu_5$. This is beyond the scope of the present analysis. 
The vector current in the magnetized medium takes the form,
\begin{align}\label{4}
j^i=&2N_c\sum_f \int \frac{d^3 \textbf{p}}{(2\pi)^3}v^i\Big(q_q f_q-q_{\bar{q}}f_{\bar{q}}\Big) \nonumber\\&=\sigma_{e} \delta^{ij} E_j +\sigma_H \epsilon^{ij}E_j,
 \end{align}
where $v_i$ is the velocity component, $\sigma_e$ and $\sigma_H$ are the electric and Hall conductivities of the magnetized QGP. Here, $\delta^{ij}$ and $\epsilon^{ij}$ are the Kronecker delta function and the antisymmetric $2 \times 2$ tensor, respectively. The first term in Eq.~(\ref{4}), which is proportional to the electric field, corresponds to the Ohmic current and the second term denotes the Hall current in the medium. 
Note that electric charge transport transverse to the magnetic field vanishes in a strongly magnetized medium, $i.e.$, $\sigma_H\sim 0$, due to $1+1-$D Landau level dynamics of the quarks/antiquarks~\cite{Kurian:2019fty}.  
Now, one needs to solve Eq.~(\ref{1}) to obtain the non-equilibrium part of the distribution function while choosing the proper collision kernel.

\subsection{RTA collision kernel}
\subsubsection{Isotropic medium}
The collisional aspects in the medium can be approximated using the RTA collision term in the Boltzmann equation as,
\begin{align}
  &C_k =-(u.p)\frac{\delta f_k}{\tau_R},  &&
    \tau_{R} =\frac{1}{5.1T\alpha_s^2 \ln (\frac{1}{\alpha_s})(1+0.12(2N_f +1))},
\end{align}
where $\tau_R$ is the thermal relaxation time of the binary scattering processes in the medium~\cite{Hosoya:1983xm,Thakur:2019bnf} and $\alpha_s$ is the one-loop coupling constant in the presence of magnetic field~\cite{Ayala:2018wux,Bandyopadhyay:2017cle}. Hence, the Boltzmann equation takes the form as follows,
 \begin{align}\label{5}
\frac{\partial f_k}{\partial t} +{\bf v}.\frac{\partial f_k}{\partial {\bf x}} +q_{f_k} [{\bf E} +{\bf v} \times {\bf B}].\frac{\partial f_k}{\partial {\bf p}} =-\frac{\delta f_k}{\tau_R}.
\end{align}
It is important to note that $1\rightarrow 2$ processes are kinematically possible and the magnetic field has a strong dependence on the collision integral in the strong magnetic field limit~\cite{Hattori:2017qih,Fukushima:2017lvb}. However, the present analysis assumes temperature as the dominant energy source in comparison with the strength of the magnetic field. This allows us to neglect the impact of the magnetic field on the thermal relaxation time of the binary processes in the medium. Employing Eq.~(\ref{2}) in Eq.~(\ref{5}), we have
\begin{widetext}
\begin{align}\label{6}
& \epsilon {\bf v}.\Big[\alpha_1 \dot{{\bf E}}+\dot{\alpha_1 }{\bf E}+\alpha_2 \ddot{{\bf E}}+\dot{\alpha_2 }\dot{{\bf E}}+\alpha_3 (\dot{{\bf E}}\times {\bf B}) +\alpha_3 ({\bf E}\times \dot{{\bf B}})+\dot{\alpha_3 }({\bf E}\times {\bf B})+\alpha_4 (\dot{{\bf E}}\times \dot{{\bf B}})+\alpha_4 (\ddot{{\bf E}} \times {\bf B}) +\dot{\alpha_4 }(\dot{{\bf E}} \times {\bf B}) \nonumber\\
&+\alpha_5 (\dot{{\bf E}}\times \dot{{\bf B}})+\alpha_5 ({\bf E}\times \ddot{{\bf B}})+\dot{\alpha_5 }({\bf E}\times \dot{{\bf B}})+\alpha_9 (\pmb{\nabla} \times {\bf \dot{B}})+\dot{\alpha_9}(\pmb{\nabla} \times {\bf B})\Big]
 +q_{f_k}{\bf v}.{\bf E}-\alpha_1 q_{f_k}{{\bf v}}.({\bf E}\times {\bf B})-\alpha_2 q_{f_k}{\bf v}.(\dot{{\bf E}}\times {\bf B})\nonumber\\
& +\alpha_3 q_{f_k}({\bf v}.{\bf E})(B^2)-\alpha_3 q_{f_k}({\bf v}.{\bf B})({\bf B}.{\bf E})+\alpha_4 q_{f_k}({\bf v}.\dot{{\bf E}})(B^2)-\alpha_4 q_{f_k}({\bf v}.{\bf B})({\bf B}.\dot{{\bf E}})+\alpha_5 q_{f_k}({\bf v}.{\bf E})(\dot{{\bf B}}.{\bf B}) -\alpha_5 q_{f_k}(\dot{{\bf B}}.{\bf v})({\bf E}.{\bf B})\nonumber\\
&-\alpha_9 q_{f_k}({\bf B.v})(\pmb{\nabla}.{\bf B})=-\frac{\epsilon}{\tau_R}\Big[\alpha_1 {\bf v}.{\bf E}+\alpha_2 {\bf v}.\dot{{\bf E}}+\alpha_3 {\bf v}.({\bf E}\times {\bf B})+\alpha_4 {\bf v}.(\dot{{\bf E}}\times {\bf B})+\alpha_5 {\bf v}.({\bf E}\times \dot{{\bf B}})+\alpha_9 {\bf v}.(\pmb{\nabla} \times {\bf B})\Big].
\end{align}
\end{widetext}
Due to the fact that the electromagnetic fields vary slowly in time to incorporate the collisional effects, we neglect the terms with second-order derivatives of the fields while estimating the electromagnetic responses in the medium. Hence, the terms with $\dot{\alpha_2}, \dot{\alpha_4}, \dot{\alpha_5}, \dot{\alpha_9}$ are negligible. By choosing the form of the magnetic field in the medium, one can determine the associated coefficients corresponding to the electric and Hall currents. Since the space-time evolution of the magnetic field in the QGP medium is not well understood, we focus the response in a constant background ${\bf B}$ to the time-dependent external electric field in the medium.
Comparing the independent terms with various tensor structures on both sides within these approximations, we obtain coupled linear differential equations for the functions $\alpha_{i}$, with $i= 1, 3$, as follows,
\begin{align}\label{7}
&  \dot{\alpha_1}=-\Big[\frac{\alpha_1}{\tau_R}+\frac{q_{f_k} \alpha_3 B^2}{\epsilon}+\frac{q_{f_k}}{\epsilon}\Big],\\
  & \dot{\alpha_3}=\frac{-\alpha_3}{\tau_R}+\frac{\alpha_1 q_{f_k}}{\epsilon}\label{7.1}.
\end{align}
Further, the $\alpha_{i}$ with $i= 2, 4$, satisfy the following coupled equations
\begin{align}\label{7.5}
&\alpha_2=-\tau_R\Big[ \alpha_1 +\frac{q_{f_k} \alpha_4 B^2}{\epsilon}\Big],
 &&{\alpha_4}=-{\tau_R}\Big[\alpha_3-\frac{\alpha_2 q_{f_k}}{\epsilon}\Big].
\end{align}
 Next, define the matrices,
 \begin{align}\label{10}
 X=\begin{pmatrix}
\alpha_1\\\alpha_3
\end{pmatrix},  
&&A=\begin{pmatrix}
 -\frac{1}{\tau_R} & -\frac{q_{f_k} B^2}{\epsilon}\\
\frac{q_{f_k}}{\epsilon} &  -\frac{1}{\tau_R}, 
\end{pmatrix},
&&& G=\begin{pmatrix}
\frac{-q_{f_k}}{\epsilon}\\ 0
\end{pmatrix},
 \end{align}
the coupled differential equations can be represented as the matrix equation as follows,
\begin{equation}\label{11}
 \frac{d X}{d t}= AX +G.   
\end{equation}
To solve Eq.~(\ref{11}), we start with the solution of the homogeneous equation $\frac{dX}{dt}=A X$ using the eigenvalues and eigenvectors of the matrix $A$. The general solution of the homogeneous part takes the following form, 
\begin{align}\label{12}
&\alpha_1 =k_1 iBe^{\big(-\frac{1}{\tau_R} +\frac{q_{f_k} iB}{\epsilon}\big)t} -k_2 iBe^{-\big(\frac{1}{\tau_R} +\frac{q_{f_k} iB}{\epsilon}\big)t},\\
&\alpha_3=k_1 e^{\big(-\frac{1}{\tau_R} +\frac{q_{f_k} iB}{\epsilon}\big)t} +k_2 e^{-\big(\frac{1}{\tau_R} +\frac{q_{f_k} iB}{\epsilon}\big)t}\label{12.2}, 
\end{align}
where $k_1$ and $k_2$ are the unknown constants, $-\frac{1}{\tau_R} \pm \frac{iq_{f_k} B }{\epsilon } $ are the eigenvalues and $\begin{pmatrix}
 iB\\ 1
\end{pmatrix}, \begin{pmatrix}
 -iB\\ 1
\end{pmatrix}$ are the eigenvectors of the matrix $A$. The general solution of the non-homogeneous equation is the sum of the solution associated homogeneous equation and a particular solution of the non-homogeneous part, say $G$, of Eq.~(\ref{11}). Employing the method of variation of constants to solve the non-homogeneous equation, the constants $k_1$ and $k_2$ are replaced with unknown functions $k_1(t)$ and $k_2(t)$. Substituting this back to Eq.~(\ref{11}) we obtain, 
\begin{align}\label{13}
&\dot{k_1} iBe^{\big(-\frac{1}{\tau_R} +\frac{q_{f_k} iB}{\epsilon}\big)t} -\dot{k_2}iBe^{-\big(\frac{1}{\tau_R} +\frac{q_{f_k} iB}{\epsilon}\big)t} =-\frac{q_{f_k}}{\epsilon},\\
&\dot{k_1} e^{\big(-\frac{1}{\tau_R} +\frac{q_{f_k} iB}{\epsilon}\big)t} +\dot{k_2}e^{-\big(\frac{1}{\tau_R} +\frac{q_{f_k} iB}{\epsilon}\big)t}=0\label{13.2}.
\end{align}
Solving the Eqs.~(\ref{13}) and~(\ref{13.2}) we have,
\begin{align}\label{14} 
&\dot{k_1}=\frac{iq_{f_k} e^{\big(\frac{1}{\tau_R} 
-\frac{q_{f_k} iB}{\epsilon}\big)t}}{2B\epsilon},
&&\dot{k_2}=-\frac{iq_{f_k} e^{\big(\frac{1}{\tau_R} +\frac{q_{f_k} iB}{\epsilon}\big)t}}{2B\epsilon}.
\end{align}
The functions $k_1(t)$ and $k_2(t)$ obtained by integrating Eq.~(\ref{14}) and  take the forms,
\begin{align}\label{15}
&k_1 =\frac{iq_{f_k} e^{\big(\frac{1}{\tau_R} -\frac{q_{f_k} iB}{\epsilon}\big)t}}{2B\epsilon \big(\frac{1}{\tau_R} -\frac{q_{f_k} iB}{\epsilon}\big)},
&&k_2 =-\frac{iq_{f_k} e^{\big(\frac{1}{\tau_R} +\frac{q_{f_k} iB}{\epsilon}\big)t}}{2B\epsilon \big(\frac{1}{\tau_R} +\frac{q_{f_k} iB}{\epsilon}\big)}.
\end{align}
Substituting Eq.~(\ref{15}) in Eq.~(\ref{12}) and Eq.~(\ref{12.2}) and employing Eq.~(\ref{7.5}), the $\alpha_i$'s take the following form,
\begin{align}\label{16}
 &\alpha_1 =-\frac{\epsilon q_{f_k}}{\tau_R [(\frac{\epsilon}{\tau_R})^2 +(q_{f_k} B)^2]},\\
 &\alpha_2 =\frac{q_{f_k} \epsilon [(\frac{\epsilon}{\tau_R})^2 -(q_{f_k} B)^2]}{[(\frac{\epsilon}{\tau_R})^2 +(q_{f_k} B)^2]^2},\\
 &\alpha_3 =-\frac{q_{f_k}^2 }{[(\frac{\epsilon}{\tau_R})^2 +(q_{f_k} B)^2]},\\
 &\alpha_4 =\frac{2q_{f_k}^2 \epsilon^2 }{\tau_R [(\frac{\epsilon}{\tau_R})^2 +(q_{f_k} B)^2]^2}\label{16.2}.
\end{align}
Note that the contribution from the term ${\bf E}\times\dot{{\bf B}}$ vanishes due to the constant background magnetic field. Employing Eqs.~(\ref{16})-(\ref{16.2}) and Eq.~(\ref{2}) in Eq.~(\ref{4}), the current density in the weakly magnetized QGP medium can be defined as,
\begin{equation}\label{17}
  {\bf j}={ j}_{e}\hat{\bf e}+{ j}_{H}\big(\hat{\bf e}\times\hat{\bf b}\big) ,
\end{equation}
where $\hat{\bf e}$ and $\hat{\bf b}$ are respectively the directions of the electric and magnetic field in the medium such that $\hat{\bf e}.\hat{\bf b}=0$. Here, ${ j}_{e}$ and ${ j}_{H}$ quantifies the electric charge transport in the direction and perpendicular to the electric field in the magnetized medium. The electric current density constitute the leading order Ohmic current ${ j}_{e}^{(0)}$ and  the additional components due to the inhomogeneity of the external electric field ${ j}_{e}^{(1)}$ as,
\begin{align}\label{18.01}
{ j}_{e}={ j}_{e}^{(0)}+{ j}_{e}^{(1)},
\end{align}
and the components take the forms as follows,
\begin{align}\label{18}
j_{e}^{(0)}=&\frac{E(t)}{3} 2N_c\sum_k \sum_f (q_{f_k})^2 \int \frac{d^3 \textbf{p}}{(2\pi)^3} p^2(-\frac{\partial f^0_k}{\partial \epsilon})\nonumber\\&\times  \frac{1}{\tau_R \big[(\frac{\epsilon}{\tau_R})^2 +(q_{f_k} B)^2\big]},
\end{align}
\begin{align}\label{19}
j_{e}^{(1)}=&\frac{\dot{E}(t)}{3}2N_c \sum_k \sum_f (q_{f_k})^2 \int \frac{d^3 \textbf{p}}{(2\pi)^3} p^2\frac{\partial f^0_k}{\partial \epsilon}\nonumber\\&\times  \frac{\big[(\frac{\epsilon}{\tau_R})^2 -(q_{f_k} B)^2\big]}{\big[(\frac{\epsilon}{\tau_R})^2 +(q_{f_k} B)^2\big]^2}.
\end{align}
Similarly, the Hall current density can be described as  
\begin{align}\label{18.02}
{ j}_{H}={ j}_{H}^{(0)}+{ j}_{H}^{(1)},
\end{align}\label{18.01}
with 
\begin{align}\label{18.3}
j_{H}^{(0)}=&\frac{E(t)}{3}2N_c \sum_k \sum_f (q_{f_k})^2 \int \frac{d^3 \textbf{p}}{(2\pi)^3} \frac{p^2}{\epsilon} (-\frac{\partial f^0_k}{\partial \epsilon}) \nonumber\\&\times \frac{q_{f_k} B}{\big[(\frac{\epsilon}{\tau_R})^2 +(q_{f_k} B)^2\big]},
\end{align}
\begin{align}\label{20}
j_{H}^{(1)}=&\frac{2\dot{E}(t)}{3}2N_c\sum_k \sum_f (q_{f_k})^2 \int \frac{d^3 \textbf{p}}{(2\pi)^3} \frac{p^2}{\epsilon} \frac{\partial f^0_k}{\partial \epsilon} \nonumber\\&\times \frac{\epsilon^2 q_{f_k} B}{\tau_R \big[(\frac{\epsilon}{\tau_R})^2 +(q_{f_k} B)^2\big]^2}.
\end{align}
Let us now proceed with the analysis of electric charge transport in an anisotropic QGP medium. 
\subsubsection{Anisotropic medium}
The success of dissipative hydrodynamics makes it reasonable to assume that the QGP created in the heavy-ion collision is slightly away from the local thermal equilibrium. In the early stages of heavy-ion
collisions, large anisotropies arise due to
the rapid longitudinal expansion of the created medium. The physics of momentum anisotropy could be understood in terms of momentum anisotropic particle distributions~\cite{Strickland:2014pga}. The anisotropic distribution function can be described in terms of isotropic distribution by re-scaling one direction in momentum space~\cite{Schenke:2006xu,Romatschke:2003ms,Schenke:2006yp}. 
\begin{align}
    f_{{\text{aniso}}}({\bf p})= \sqrt{1+\xi}\,f_{{\text{iso}}}\Big(\sqrt{p^2+\xi({\bf p}\cdot{\bf n})^2}\Big),
\end{align}
where $f_{{\text{iso}}}=f^0$ is the isotropic distribution function and $\xi=\frac{\langle {\bf p}^2_T\rangle}{2\langle p_L^2\rangle}-1$ is the anisotropic parameter such that ${\bf p}_T={\bf p}-{\bf n}({\bf p}\cdot{\bf n})$ and $p_L={\bf p}\cdot{\bf n}$ with  ${\bf n}$ is the direction of anisotropy.  The current focus is on a weakly anisotropic medium with $\xi\ll 1$ and we have~\cite{Srivastava:2015via},
\begin{align}
    f_{{\text{aniso}}}({\bf p})= f^0-\frac{\xi}{2\epsilon T}({\bf p}\cdot{\bf n})^2 (f^0)^2\exp{\Big(\frac{\epsilon \mp \mu}{T}\Big)},
\end{align}
where ${\bf p}=(p\sin\theta \cos\phi,\, p\sin\theta \sin\phi,\, p\cos\theta)$ and ${\bf n}=(\cos\alpha,\, 0,\,  \sin\alpha)$. Solving the Boltzmann equation in the magnetized medium and following the same formalism in Ref.~\cite{Srivastava:2015via}, we obtain the electric current density in the anisotropic medium as follows,
\begin{align}
    (j_{e})_{{\text{aniso}}} =j_e^{(0)} +\delta j_e^{(0)} +j_e^{(1)} +\delta j_e^{(1)},
\end{align}
where the isotropic components $j_e^{(0)}$ and $j_e^{(1)}$ are defined in Eq.~(\ref{18}) and Eq.~(\ref{19}), respectively. The terms $\delta j_e^{(0)}$ and $\delta j_e^{(1)}$ quantify the effect of anisotropy on the electric current density in the medium and take the forms as follows,
\begin{align}\label{2.18}
\delta j_{e}^{(0)}=&-\xi\frac{E(t)}{3} N_c\sum_k \sum_f (q_{f_k})^2 \frac{1}{6\pi^2 T^2}\int dp \frac{ p^6}{\epsilon} (f^0_k)^2\nonumber\\&\times \exp{\Big(\frac{\epsilon \mp \mu}{T}\Big)} \frac{1}{\tau_R \big[(\frac{\epsilon}{\tau_R})^2 +(q_{f_k} B)^2\big]},
\end{align}
\begin{align}\label{2.19}
\delta j_{e}^{(1)}=&\xi\frac{\dot{E}(t)}{3}N_c \sum_k \sum_f (q_{f_k})^2 \frac{1}{6\pi^2 T^2} \int dp \frac{ p^6}{\epsilon} (f^0_k)^2\nonumber\\&\times \exp{\Big(\frac{\epsilon \mp \mu}{T}\Big)} \frac{\big[(\frac{\epsilon}{\tau_R})^2 -(q_{f_k} B)^2\big]}{\big[(\frac{\epsilon}{\tau_R})^2 +(q_{f_k} B)^2\big]^2}.
\end{align}
Notably, Eq.~(\ref{2.18}) reduces back to the results of~\cite{Srivastava:2015via} in the case of vanishing magnetic field. Similarly, the Hall current density in the anisotropic medium is defined as,  
\begin{align}
 (j_{H})_{{\text{aniso}}} =j_H^{(0)} +\delta j_H^{(0)} +j_H^{(1)} +\delta j_H^{(1)},
\end{align}
with anisotropic contributions to the Hall current take the following forms,
\begin{align}\label{2.18.3}
\delta j_{H}^{(0)}=&-\xi\frac{E(t)}{3}N_c \sum_k \sum_f (q_{f_k})^2 \frac{1}{6\pi^2 T^2} \int dp \frac{ p^6}{\epsilon^2}  (f^0_k)^2 \nonumber\\&\times \exp{\Big(\frac{\epsilon \mp \mu}{T}\Big)} \frac{q_{f_k} B}{\big[(\frac{\epsilon}{\tau_R})^2 +(q_{f_k} B)^2\big]},
\end{align}
\begin{align}\label{2.20}
\delta j_{H}^{(1)}=&\xi\frac{2\dot{E}(t)}{3}N_c\sum_k \sum_f (q_{f_k})^2 \frac{1}{6\pi^2 T^2} \int dp \frac{ p^6}{\epsilon^2} (f^0_k)^2 \nonumber\\&\times \exp{\Big(\frac{\epsilon \mp \mu}{T}\Big)} \frac{\epsilon^2 q_{f_k} B}{\tau_R \big[(\frac{\epsilon}{\tau_R})^2 +(q_{f_k} B)^2\big]^2}.
\end{align}
The isotropic terms $j_H^{(0)}$ and $j_H^{(1)}$ are described in Eq.~(\ref{18.3}) and Eq.~(\ref{20}). 

\subsection{BGK collision kernel}
The collisional aspects of hot QCD medium can be described using the BGK collision term~\cite{Schenke:2006xu,Kumar:2017bja,Khan:2020rdw}. The advantage of the BGK collision term over conventional RTA kernel is it naturally preserves number conservation, $i.e.$, \begin{equation}
 \int \frac{d^3 \textbf{p}}{(2\pi)^3} C(f_k) =0.   
\end{equation}
We closely follow Refs.~\cite{Jiang:2016dkf,Carrington:2003je} to incorporate the BGK collisional aspects in the estimation of current density in the weakly magnetized medium. The BGK collision kernel can be defined as,
\begin{equation}\label{21}
C(f_k)=-\nu \Big[f_k-\frac{N}{N^0}f^0_k\Big]\\
 =-\nu \Big[\delta f_k -\frac{f^0_k}{N^0} \int \frac{d^3 {\bf p}}{(2\pi)^3} \delta f_k\Big],
\end{equation}
where, 
\begin{align}\label{22}
&N= \sum_k\int \frac{d^3 {\bf p}}{(2\pi)^3}  f_k(p), && N^0 =\sum_k\int \frac{d^3 {\bf p}}{(2\pi)^3}  f^0_k (p),
\end{align}
are the particle number densities. Here, the $\nu$ denotes the collisional frequency, which acts as the input parameter of the transport process in the medium and is independent of the particle momentum. In the present analysis, $\nu$ is fixed as the thermal average of inverse of thermal relaxation time for the binary collisions in the medium. Note that the BGK kernel reduces back to RTA term in the limit $\frac{N}{N^0}=1$. 

Solving the Boltzmann Eq.~(\ref{5}) within BGK kernel, we obtain the $\delta f_k$ as,
\begin{equation}\label{23}
\delta f_k =\delta f_k^{(0)} +\nu D^{-1}\frac{f_k^0}{N^0}\eta \frac{1}{1-\lambda},
\end{equation}
where the RTA equivalent non-equilibrium part of the distribution function $\delta f_k^{(0)}$ can be defined as,
\begin{align}\label{24}
  \delta f_k^{(0)} =-q_{f_k}(\textbf{E} +\textbf{v} \times \textbf{B}).\frac{\partial f^0_k}{\partial \textbf{p}} D^{-1}, 
\end{align}
with $D=\partial_0 +\nu$. The functions $\eta$ and $\lambda$ take  the forms as follows,
\begin{align}\label{25}
&\eta =\int\frac{d^3 \textbf{p}^{'}}{(2\pi)^3} \delta f_k^{(0)},
&&\lambda =\frac{i\nu }{N_0} \int\frac{d^3\textbf{p}^{''}}{(2\pi)^3} f_k^0 D^{-1}.
\end{align}
\begin{figure*}
    \centering
    \centering
    \hspace{-1.5cm}
    \includegraphics[width=0.515\textwidth]{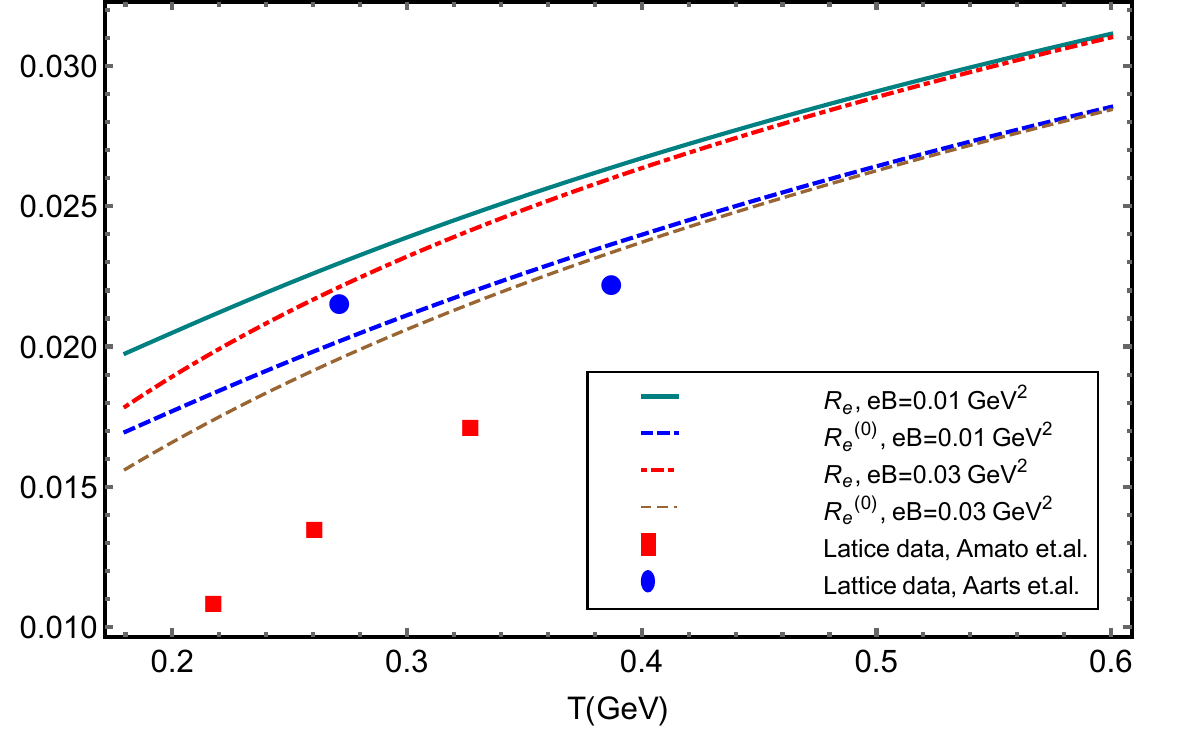}
    \hspace{-.5cm}
    \includegraphics[width=0.575\textwidth]{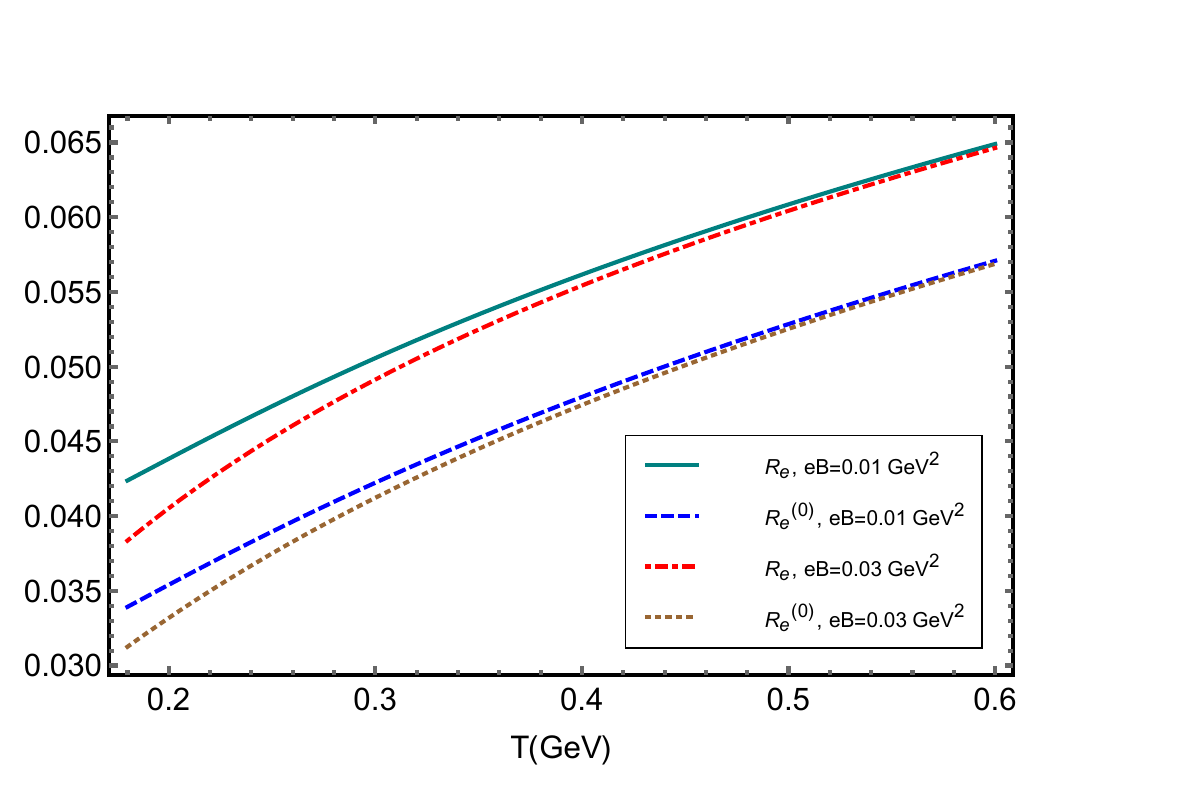}
    \hspace{-2.5cm}
    \caption{Temperature dependence of $R_{e}^{(0)}$ and $R_{e}$ at $\mid eB\mid=0.01$ GeV$^2$ and $0.03$ GeV$^2$ with $\mu=100$ MeV and $\tau_E =5$ fm within RTA (left panel) and BGK (right panel) collision kernels. The RTA results are compared with lattice data from Refs.~\cite{Amato:2013naa} and~\cite{Aarts:2007wj}.}
\label{f1}
\end{figure*}
Expanding $D^{-1}$ and $\lambda$ in leading order in the presence of inhomogeneous fields, we obtain $\delta f_k$ as,
\begin{align}\label{26}
\delta f_k=(\textbf{p}.\mathbf{\Xi}) \frac{\partial f_k^0}{\partial \epsilon} +\frac{\nu f_k^0}{N^0} \int \frac{d^3 {\bf p}}{(2\pi)^3}( \textbf{p}. \mathbf{\Theta}) \frac{\partial f^0_k}{\partial \epsilon},
\end{align}
where, 
\begin{align}\label{27}
\mathbf{\Theta} &=\frac{1}{\nu} \bigg[\alpha_1 \textbf{E} +\big(\alpha_2-\nu^{-1} \alpha_1\big)\dot{\textbf{E}} +\alpha_3 \big(\textbf{E} \times \textbf{B}\big)+\big(\alpha_4-\nu^{-1} \alpha_3\big)\nonumber\\& \big(\dot{\textbf{E}} \times\textbf{B}\big) +\big(\alpha_5-\nu^{-1} \alpha_3\big)\big(\textbf{E} \times \dot{\textbf{B}}\big) -\nu^{-1} \alpha_4 \big(\ddot{\textbf{E}} \times \textbf{B}\big)\nonumber\\& -\nu^{-1} \big(\alpha_4 +\alpha_5 \big)\big(\dot{\textbf{E}} \times \dot{\textbf{B}}\big) -\nu^{-1} \alpha_5 \big(\textbf{E} \times \ddot{\textbf{B}}\big)\bigg].
\end{align}
Employing Eq.~(\ref{26}) in Eq.~(\ref{4}), we obtain the leading order electrical and Hall current density respectively as,
\begin{widetext}
\begin{align}\label{28}
&j_{e}^{(0)}=\frac{E(t)}{3}2N_c \sum_k \sum_f (q_{f_k})^2 \int \frac{d^3 \textbf{p}}{(2\pi)^3}p \Bigg\{ (-\frac{\partial f^0_k}{\partial \epsilon}) \bigg[\frac{ \epsilon \nu}{(\epsilon \nu)^2 +(q_{f_k} B)^2}\bigg]+ \frac{f^0_k}{N^0 \epsilon} \int \frac{d^3 {\bf p^{'}}}{(2\pi)^3} p^{'} \bigg[\frac{ \epsilon^{'} \nu}{(\epsilon^{'} \nu)^2 +(q_{f_k} B)^2}\bigg](-\frac{\partial f^0_k}{\partial \epsilon^{'}}) \Bigg\},
\end{align}
\begin{align}
&j_{e}^{(1)}=\frac{\dot{E}(t)}{3}2N_c \sum_k \sum_f (q_{f_k})^2 \int \frac{d^3 \textbf{p}}{(2\pi)^3} p \Bigg\{ \frac{\partial f^0_k}{\partial \epsilon}\bigg[ \frac{\epsilon [(\epsilon \nu)^2 -(q_{f_k} B)^2]}{[(\epsilon \nu)^2 +(q_{f_k} B)^2]^2}\bigg] +\frac{f^0_k}{N^0 \epsilon} \int \frac{d^3 {\bf p^{'}}}{(2\pi)^3} p^{'} \bigg[\frac{ 2\epsilon^{'3} \nu^2}{[(\epsilon^{'} \nu)^2 +(q_{f_k} B)^2 ]^2}\bigg]\frac{\partial f^0_k}{\partial \epsilon^{'}} \Bigg\},
\end{align}
\begin{align}
&j_{H}^{(0)}=\frac{E(t)}{3}2N_c \sum_k \sum_f (q_{f_k})^2 \int \frac{d^3 \textbf{p}}{(2\pi)^3} \frac{p^2}{\epsilon} \Bigg\{ (-\frac{\partial f^0_k}{\partial \epsilon}) \bigg[\frac{q_{f_k} B}{(\epsilon \nu)^2 +(q_{f_k} B)^2}\bigg]+ \frac{f^0_k}{N^0} \int \frac{d^3 {\bf p^{'}}}{(2\pi)^3 }p^{'} \bigg[\frac{q_{f_k} B}{(\epsilon^{'} \nu)^2 +(q_{f_k} B)^2}\bigg] (-\frac{\partial f^0_k}{\partial \epsilon^{'}}) \Bigg\},\label{30} 
\end{align}
\begin{align}
&j_{H}^{(1)}=\frac{\dot{E}(t)}{3}2N_c\sum_k \sum_f (q_{f_k})^2 \int \frac{d^3 \textbf{p}}{(2\pi)^3} \frac{p^2}{\epsilon} \Bigg\{ \frac{\partial f^0_k}{\partial \epsilon}\bigg[\frac{2q_{f_k} B \epsilon^2 \nu}{ [(\epsilon \nu)^2 +(q_{f_k} B)^2]^2} \bigg] + \frac{f^0_k}{N^0} \int \frac{d^3 {\bf p^{'}}}{(2\pi)^3}p^{'}\frac{\epsilon^{'2} \nu [3+(\frac{q_{f_k} B}{\nu})^2]}{[(\epsilon^{'} \nu)^2 +(q_{f_k} B)^2]^2} \frac{\partial f^0_k}{\partial \epsilon^{'}} \Bigg\}.\label{31}
\end{align}
\end{widetext}
\section{Results and Discussions}
\begin{figure}[h]
\hspace{-.9cm}
\includegraphics[scale=.515]{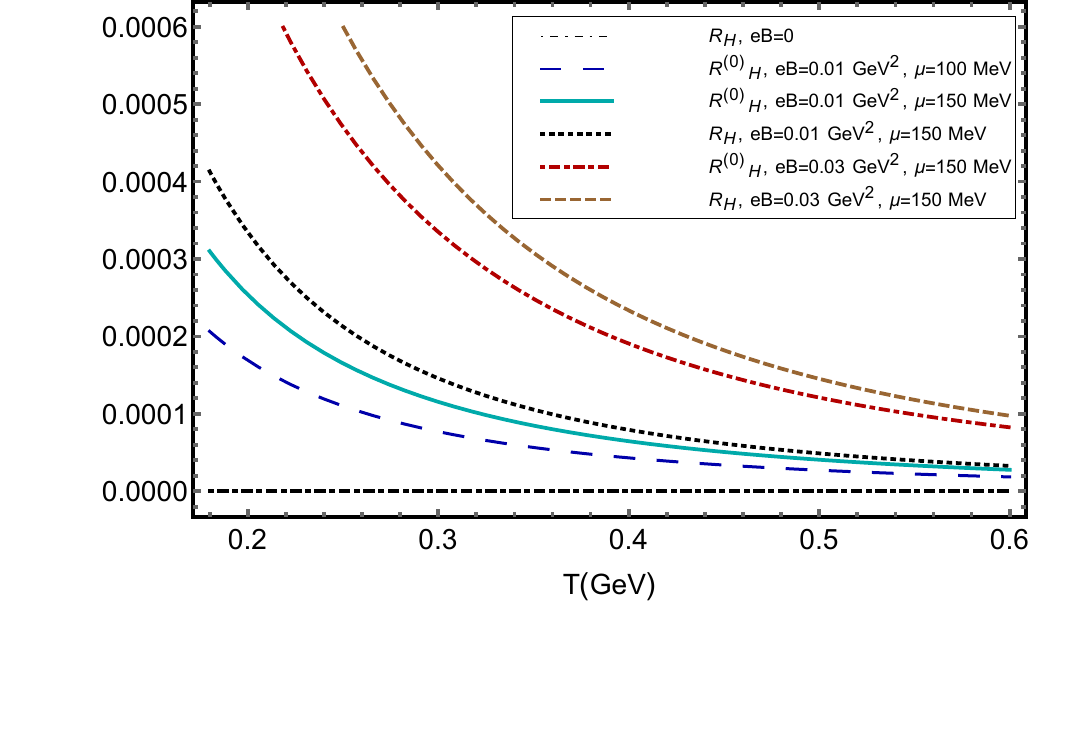} 
\caption{Dependence of chemical potential and strength of the magnetic field on the temperature behaviour of Hall current and the additional component of Hall current due to the inhomogeneity of the external electric field.}
\label{f2}
\end{figure}
\begin{figure*}
    \centering
   \hspace{-1.25cm}
    \includegraphics[width=0.521\textwidth]{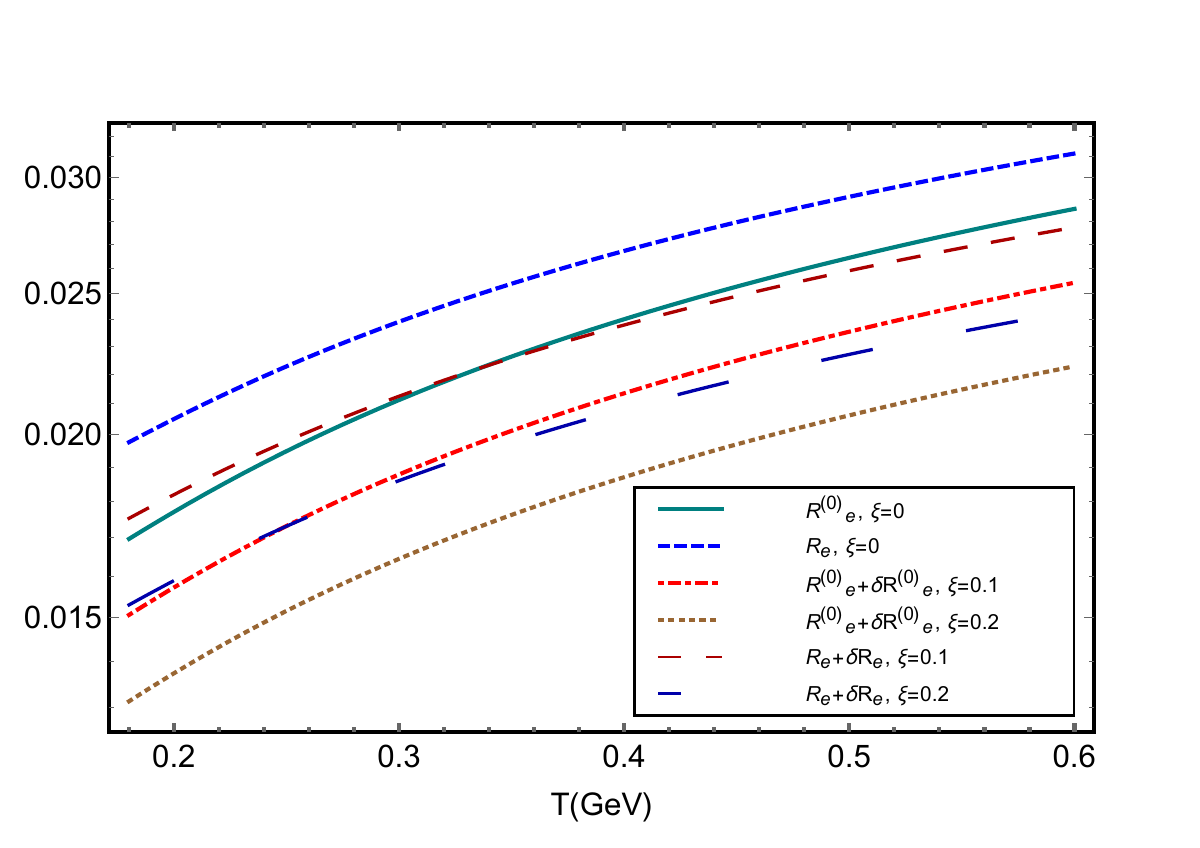}
    \includegraphics[width=0.521\textwidth]{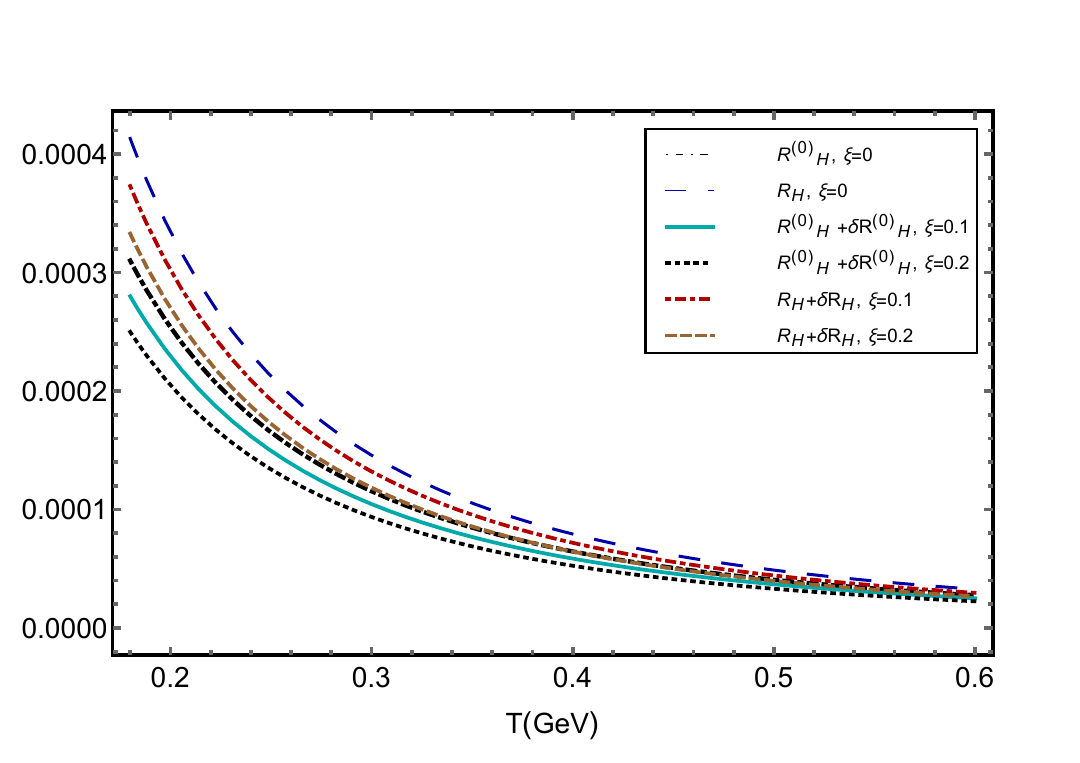}
    \hspace{-.65cm}
    \caption{Effect of momentum anisotropy on the temperature dependence of electric charge transport (left panel) and Hall current (right panel) within RTA at $\mid eB\mid=0.01$ GeV$^2$ and $\mu=150$ MeV.}
\label{f2.1}
\end{figure*}
We initiate the discussions with the temperature dependence of the response of the magnetized QGP to the inhomogeneous electric field. In the weakly magnetized medium, the electrical and Hall conductivities quantify the system response along the direction of the electric field (Ohmic) and transverse to the direction of both electric and magnetic fields (Hall), respectively. To study the temperature dependence of the Ohmic and Hall conductivities along with the additional components due to the inhomogeneity of the electric field in the weakly magnetized medium, we define the ratios,
\begin{align}\label{32}
  R_{e/H} =R_{e/H}^{(0)}+R_{e/H}^{(1)},
\end{align}
with
\begin{align}\label{33}
  &R_{e/H}^{(0)} =\frac{j_{e/H}^{(0)}}{ET}, &&R_{e/H}^{(1)} =\frac{j_{e/H}^{(1)}}{ET}.
\end{align}
The ratio $R_{e/H}^{(0)}$ denotes the conductivities in the limit of constant electromagnetic fields, $i.e.$, $R_{e}^{(0)}=\sigma_{e}$ and $R_{H}^{(0)}=\sigma_{H}$. The space-time profile of the electric field $E$ in the medium to quantify the effect of inhomogeneity in the current densities has been chosen as~\cite{Hongo:2013cqa,Satow:2014lia}, 
\begin{align}
 e{\bf E_y} = \hat{y}\,  eE_0 \frac{b}{2R} \exp\Big(-\frac{x^2}{2\sigma_x^2}-\frac{y^2}{2\sigma_y^2}-\frac{\eta_s^2}{2\sigma^2_{\eta}}-\frac{\tau}{\tau_E}\Big),   
\end{align}
where $\sigma_x$, $\sigma_y$, $\sigma_{\eta}$ determine the spatial width of the field. Here, $\tau_{E}$ is the duration time (lifetime) of the electric field, $R=6.38$ fm is the radius of the nucleus and $b$ is the impact parameter. The strength of the inhomogeneity of the electric field in time can be quantified in terms of $\tau_E$ ($\frac{\dot{\bf E}}{{\bf E}}\propto -\frac{1}{\tau_E}$). 

The temperature dependence the ratios  $R_{e}^{(0)}$ and  $R_{e}$ is depicted in Fig.~\ref{f1} (left panel) at $\mid eB\mid=0.01$ GeV$^2$ and $0.03$ GeV$^2$. We observed that the effect of inhomogeneity of the field has a significant impact on the electromagnetic response of the medium. The additional component of the current density $j_{e}^{(1)}$ described in Eq.~(\ref{19}) is higher order in $\tau_R$ in comparison to the leading order current density, $j_{e}^{(0)}$. This observation is in line with the results of Ref.~\cite{Satow:2014lia}. The present analysis is on the weakly magnetized QGP in which the strength of the magnetic field is subdominant in comparison to the temperature scale of the medium. The magnetic field dependence on the current density is entering through the Lorentz force term in the relativistic Boltzmann equation and is more prominent in the lower temperature regimes near the transition temperature $T_c$.

The collisional aspects in the estimation of the electric current density are incorporated through the RTA and BGK collision kernels. The effect of the collisions in the temperature dependence of electric charge transport is plotted in Fig.~\ref{f1} (right panel). The results of electric charge response within the BGK collision term show qualitatively similar behaviour as the RTA results with a significant shift throughout the temperature regime under consideration. 
This observation is consistent with the temperature behaviour of longitudinal electrical conductivity in the presence of a strong magnetic field within the BGK collision kernel, as discussed in Ref.~\cite{Khan:2020rdw}. The first term in Eqs.~(\ref{28})-(\ref{31}) describe the RTA results and other terms give further corrections to the current densities in the magnetized medium. We observe that the collisional effects are critically depending on the strength of the magnetic field and temperature of the medium. 

The Lorenz force results in the Hall current in the direction transverse to the particle velocity and the magnetic field in the medium. From Eq.~(\ref{18}) and Eq.~(\ref{20}), we can understand the Hall current is higher order in $\tau_R$ in comparison to the Ohmic current in the medium. The dependence of Hall current on the strength of the magnetic field, inhomogeneity of the electric field, and quark chemical potential are depicted in Fig.~\ref{f2}. We observe that the Hall current is subdominant compared to Ohmic current and vanishes at the limit $\mu=0$. The Hall current varies with the strength of the magnetic field as it is proportional to the factor $\frac{q_{f_k}B}{(\frac{\epsilon}{\tau_R})^2 +(q_{f_k} B)^2}$ within RTA. However, in the strong magnetic field limit, the Hall current vanishes due to the dimensionally reduced motion of quarks and antiquarks in the direction of the field. We also observed that the effect of inhomogeneity of the electric field on the Hall current is more pronounced in lower temperature regime and is critically depending on the quark chemical potential. The effect of collisions in the QGP medium on the Hall current and the additional component of the Hall current due to the inhomogeneity of the field are described in Eqs.~(\ref{18.3}),~(\ref{20}),~(\ref{30}) and~(\ref{31}).  

\begin{figure}[h]
\centering
\hspace{-1.5cm}
\includegraphics[scale=.5]{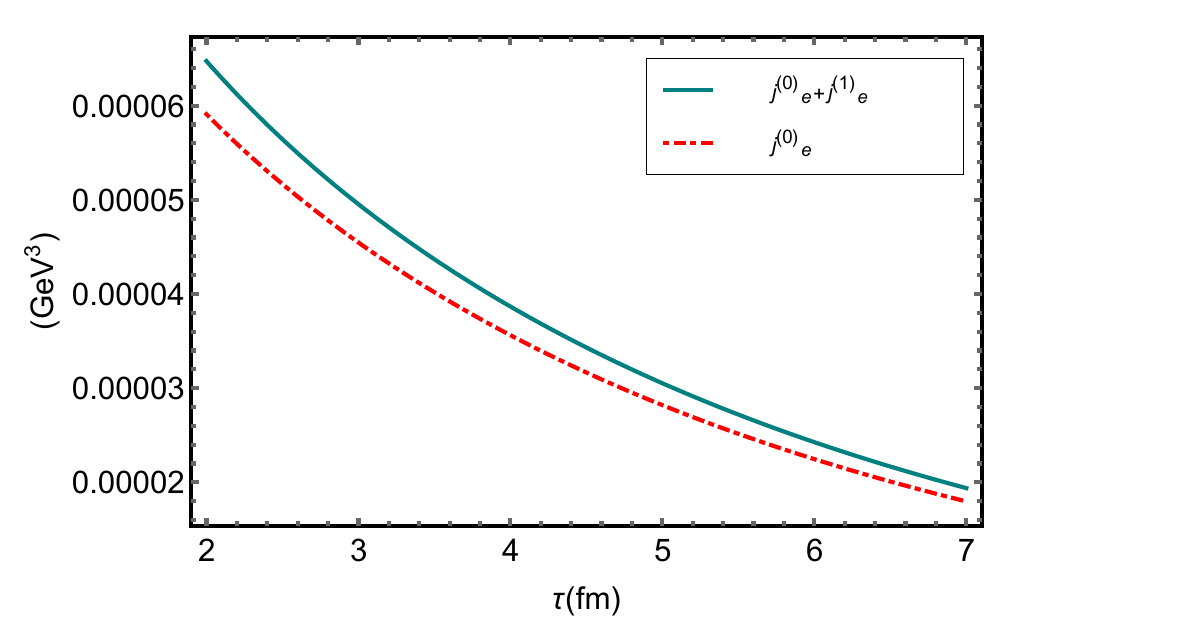} 
\caption{Proper time ($\tau$) evolution of the electric charge transport (Ohmic) in the QGP medium at $B=0$.}
\label{f3}
\end{figure}

The effect of momentum anisotropy to the electric charge transport can be described in terms of the ratios as,
\begin{align}\label{01.32}
  \delta R_{e/H} =\delta R_{e/H}^{(0)}+\delta R_{e/H}^{(1)},
\end{align}
where,
\begin{align}\label{01.33}
  &\delta R_{e/H}^{(0)} =\frac{\delta j_{e/H}^{(0)}}{ET}, &&\delta R_{e/H}^{(1)} =\frac{\delta j_{e/H}^{(1)}}{ET}.
\end{align}
The anisotropic corrections to the electric and Hall current densities are described in Eq.~(\ref{2.18})-Eq.~(\ref{2.20}). The Eq.~(\ref{01.32}) gives correction to the electric and Hall conductivities as described in  Eq.~(\ref{32}) in the anisotropic medium. The effect of anisotropy on the electric charge transport is depicted in Fig.~\ref{f2.1}. We observe that the ratio $R^0_{e}$ and $R_{e}$ decreases with the increase in the strength of anisotropy. This observation is consistent with the results of Ref.~\cite{Srivastava:2015via} at ${\bf B}=0$. The same observation holds true for the Hall current density in the presence of inhomogeneous fields.

The effect of inhomogeneity of the external electric field to electric charge transport is quantified in the case of boost-invariant one-dimensional expansion~\cite{Bjorken:1982qr} of the medium at $B=0$. The proper time behaviour of current is plotted in Fig.~\ref{f3}. The response of the medium to the decaying electric field is more visible in the initial stages of heavy-ion collision. It is also observed that the time evolution of electric field that generates additional component of the current density $j_{e}^{(1)}$ has a significant effect on the response of the system.

\section{Conclusion and Outlook}
In this article, we have studied the electric charge transport in the magnetized hot QCD/QGP medium in the presence of an external inhomogeneous electric field at finite quark chemical potential. The magnetic field is assumed to be weak in comparison to the temperature energy scale in the system. The magnetic field introduces anisotropy in the charge transport in the medium. We have investigated the effect of the inhomogeneity of the electric field to the medium's response in the direction of the electric field and transverse to the direction of both the electric and magnetic fields, respectively, for an isotropic and anisotropic medium. We have considered the case where the inhomogeneity of the field is small in which collisional aspects are significant in the transport process. We have obtained the non-equilibrium correction to the momentum distribution of quarks/antiquarks and gluons by solving the Boltzmann equation in the presence of an inhomogeneous electric field with a proper collision kernel. 

The electromagnetic responses have been studied within RTA and BGK collision kernels in the weakly magnetized QGP medium. We found that the inhomogeneous effects to the response of the medium depend on the strength of the magnetic field and the collisional aspects of the QGP. We have investigated the dependence of quark chemical potential and strength of the magnetic field on the temperature dependence of Hall conductivity and the additional component due to the inhomogeneity of the external electric field. We have also investigated the response of the QGP to the inhomogeneous electric field in the case of boost-invariant expansion of the medium in the absence of a magnetic field. The results showed that the inhomogeneous effects of the electric field on the electric charge transport in the weakly magnetized medium are non-negligible both in the direction of the electric field and in the direction perpendicular to the electric and magnetic fields in the collisional medium. Further, we have studied the effect of momentum anisotropy on the medium response within the RTA. It is observed that the momentum anisotropy has a visible impact on both electric and Hall current densities in the magnetized medium. These effects may perhaps have a significant impact on the charge-dependent directed flow of final particles in the heavy-ion collision as it is sensitive to the induced current in the medium~\cite{Hirono:2012rt}.  Further, it is important to incorporate the back reaction of the medium to electromagnetic fields for the realistic magnetohydrodynamical formulation of the created matter in the collision experiments. These aspects will be taken in a follow-up work. 

An immediate extension of the present analysis is to investigate the interplay of effects of hot QCD equation of state and inhomogeneity of the fields in the electromagnetic responses by effectively modeling the weakly magnetized medium. The present analysis of electric charge transport in the magnetized medium is the first step in this direction. We intend to investigate the electric charge transport in the regime where the effects of inhomogeneity of the field are large in comparison with the collisional effects of the medium, in the near future. 
The magnetic field will also introduce anisotropy in momentum and thermal transport in the medium. The investigation of all components of viscous coefficients and thermal conductivity in a weakly magnetized medium within the effective kinetic theory would be another interesting direction to work in the near future.

\section{Acknowledgments}
M.K. would like to acknowledge the Indian Institute of Technology Gandhinagar for Institute postdoctoral fellowship. The authors record a deep sense of gratitude to the people of India for their generous support for research in fundamental sciences.


\bibliography{ref}{}

\begin{thebibliography}{75}%
\makeatletter
\providecommand \@ifxundefined [1]{%
 \@ifx{#1\undefined}
}%
\providecommand \@ifnum [1]{%
 \ifnum #1\expandafter \@firstoftwo
 \else \expandafter \@secondoftwo
 \fi
}%
\providecommand \@ifx [1]{%
 \ifx #1\expandafter \@firstoftwo
 \else \expandafter \@secondoftwo
 \fi
}%
\providecommand \natexlab [1]{#1}%
\providecommand \enquote  [1]{``#1''}%
\providecommand \bibnamefont  [1]{#1}%
\providecommand \bibfnamefont [1]{#1}%
\providecommand \citenamefont [1]{#1}%
\providecommand \href@noop [0]{\@secondoftwo}%
\providecommand \href [0]{\begingroup \@sanitize@url \@href}%
\providecommand \@href[1]{\@@startlink{#1}\@@href}%
\providecommand \@@href[1]{\endgroup#1\@@endlink}%
\providecommand \@sanitize@url [0]{\catcode `\\12\catcode `\$12\catcode
  `\&12\catcode `\#12\catcode `\^12\catcode `\_12\catcode `\%12\relax}%
\providecommand \@@startlink[1]{}%
\providecommand \@@endlink[0]{}%
\providecommand \url  [0]{\begingroup\@sanitize@url \@url }%
\providecommand \@url [1]{\endgroup\@href {#1}{\urlprefix }}%
\providecommand \urlprefix  [0]{URL }%
\providecommand \Eprint [0]{\href }%
\providecommand \doibase [0]{http://dx.doi.org/}%
\providecommand \selectlanguage [0]{\@gobble}%
\providecommand \bibinfo  [0]{\@secondoftwo}%
\providecommand \bibfield  [0]{\@secondoftwo}%
\providecommand \translation [1]{[#1]}%
\providecommand \BibitemOpen [0]{}%
\providecommand \bibitemStop [0]{}%
\providecommand \bibitemNoStop [0]{.\EOS\space}%
\providecommand \EOS [0]{\spacefactor3000\relax}%
\providecommand \BibitemShut  [1]{\csname bibitem#1\endcsname}%
\let\auto@bib@innerbib\@empty
\bibitem [{\citenamefont {Acharya}\ \emph {et~al.}(2020)\citenamefont {Acharya}
  \emph {et~al.}}]{Acharya:2019ijj}%
  \BibitemOpen
  \bibfield  {author} {\bibinfo {author} {\bibfnamefont {S.}~\bibnamefont
  {Acharya}} \emph {et~al.} (\bibinfo {collaboration} {ALICE}),\ }\href
  {\doibase 10.1103/PhysRevLett.125.022301} {\bibfield  {journal} {\bibinfo
  {journal} {Phys. Rev. Lett.}\ }\textbf {\bibinfo {volume} {125}},\ \bibinfo
  {pages} {022301} (\bibinfo {year} {2020})},\ \Eprint
  {http://arxiv.org/abs/1910.14406} {arXiv:1910.14406 [nucl-ex]} \BibitemShut
  {NoStop}%
\bibitem [{\citenamefont {Adam}\ \emph {et~al.}(2019)\citenamefont {Adam} \emph
  {et~al.}}]{Adam:2019wnk}%
  \BibitemOpen
  \bibfield  {author} {\bibinfo {author} {\bibfnamefont {J.}~\bibnamefont
  {Adam}} \emph {et~al.} (\bibinfo {collaboration} {STAR}),\ }\href {\doibase
  10.1103/PhysRevLett.123.162301} {\bibfield  {journal} {\bibinfo  {journal}
  {Phys. Rev. Lett.}\ }\textbf {\bibinfo {volume} {123}},\ \bibinfo {pages}
  {162301} (\bibinfo {year} {2019})},\ \Eprint
  {http://arxiv.org/abs/1905.02052} {arXiv:1905.02052 [nucl-ex]} \BibitemShut
  {NoStop}%
\bibitem [{\citenamefont {Skokov}\ \emph {et~al.}(2009)\citenamefont {Skokov},
  \citenamefont {Illarionov},\ and\ \citenamefont {Toneev}}]{Skokov:2009qp}%
  \BibitemOpen
  \bibfield  {author} {\bibinfo {author} {\bibfnamefont {V.}~\bibnamefont
  {Skokov}}, \bibinfo {author} {\bibfnamefont {A.}~\bibnamefont {Illarionov}},
  \ and\ \bibinfo {author} {\bibfnamefont {V.}~\bibnamefont {Toneev}},\ }\href
  {\doibase 10.1142/S0217751X09047570} {\bibfield  {journal} {\bibinfo
  {journal} {Int. J. Mod. Phys. A}\ }\textbf {\bibinfo {volume} {24}},\
  \bibinfo {pages} {5925} (\bibinfo {year} {2009})},\ \Eprint
  {http://arxiv.org/abs/0907.1396} {arXiv:0907.1396 [nucl-th]} \BibitemShut
  {NoStop}%
\bibitem [{\citenamefont {Zhong}\ \emph {et~al.}(2014)\citenamefont {Zhong},
  \citenamefont {Yang}, \citenamefont {Cai},\ and\ \citenamefont
  {Feng}}]{Zhong:2014cda}%
  \BibitemOpen
  \bibfield  {author} {\bibinfo {author} {\bibfnamefont {Y.}~\bibnamefont
  {Zhong}}, \bibinfo {author} {\bibfnamefont {C.-B.}\ \bibnamefont {Yang}},
  \bibinfo {author} {\bibfnamefont {X.}~\bibnamefont {Cai}}, \ and\ \bibinfo
  {author} {\bibfnamefont {S.-Q.}\ \bibnamefont {Feng}},\ }\href {\doibase
  10.1155/2014/193039} {\bibfield  {journal} {\bibinfo  {journal} {Adv. High
  Energy Phys.}\ }\textbf {\bibinfo {volume} {2014}},\ \bibinfo {pages}
  {193039} (\bibinfo {year} {2014})},\ \Eprint {http://arxiv.org/abs/1408.5694}
  {arXiv:1408.5694 [hep-ph]} \BibitemShut {NoStop}%
\bibitem [{\citenamefont {Voronyuk}\ \emph {et~al.}(2011)\citenamefont
  {Voronyuk}, \citenamefont {Toneev}, \citenamefont {Cassing}, \citenamefont
  {Bratkovskaya}, \citenamefont {Konchakovski},\ and\ \citenamefont
  {Voloshin}}]{Voronyuk:2011jd}%
  \BibitemOpen
  \bibfield  {author} {\bibinfo {author} {\bibfnamefont {V.}~\bibnamefont
  {Voronyuk}}, \bibinfo {author} {\bibfnamefont {V.}~\bibnamefont {Toneev}},
  \bibinfo {author} {\bibfnamefont {W.}~\bibnamefont {Cassing}}, \bibinfo
  {author} {\bibfnamefont {E.}~\bibnamefont {Bratkovskaya}}, \bibinfo {author}
  {\bibfnamefont {V.}~\bibnamefont {Konchakovski}}, \ and\ \bibinfo {author}
  {\bibfnamefont {S.}~\bibnamefont {Voloshin}},\ }\href {\doibase
  10.1103/PhysRevC.83.054911} {\bibfield  {journal} {\bibinfo  {journal} {Phys.
  Rev. C}\ }\textbf {\bibinfo {volume} {83}},\ \bibinfo {pages} {054911}
  (\bibinfo {year} {2011})},\ \Eprint {http://arxiv.org/abs/1103.4239}
  {arXiv:1103.4239 [nucl-th]} \BibitemShut {NoStop}%
\bibitem [{\citenamefont {Deng}\ and\ \citenamefont
  {Huang}(2012)}]{Deng:2012pc}%
  \BibitemOpen
  \bibfield  {author} {\bibinfo {author} {\bibfnamefont {W.-T.}\ \bibnamefont
  {Deng}}\ and\ \bibinfo {author} {\bibfnamefont {X.-G.}\ \bibnamefont
  {Huang}},\ }\href {\doibase 10.1103/PhysRevC.85.044907} {\bibfield  {journal}
  {\bibinfo  {journal} {Phys. Rev. C}\ }\textbf {\bibinfo {volume} {85}},\
  \bibinfo {pages} {044907} (\bibinfo {year} {2012})},\ \Eprint
  {http://arxiv.org/abs/1201.5108} {arXiv:1201.5108 [nucl-th]} \BibitemShut
  {NoStop}%
\bibitem [{\citenamefont {Fukushima}\ \emph {et~al.}(2008)\citenamefont
  {Fukushima}, \citenamefont {Kharzeev},\ and\ \citenamefont
  {Warringa}}]{Fukushima:2008xe}%
  \BibitemOpen
  \bibfield  {author} {\bibinfo {author} {\bibfnamefont {K.}~\bibnamefont
  {Fukushima}}, \bibinfo {author} {\bibfnamefont {D.~E.}\ \bibnamefont
  {Kharzeev}}, \ and\ \bibinfo {author} {\bibfnamefont {H.~J.}\ \bibnamefont
  {Warringa}},\ }\href {\doibase 10.1103/PhysRevD.78.074033} {\bibfield
  {journal} {\bibinfo  {journal} {Phys. Rev. D}\ }\textbf {\bibinfo {volume}
  {78}},\ \bibinfo {pages} {074033} (\bibinfo {year} {2008})},\ \Eprint
  {http://arxiv.org/abs/0808.3382} {arXiv:0808.3382 [hep-ph]} \BibitemShut
  {NoStop}%
\bibitem [{\citenamefont {Kharzeev}\ \emph {et~al.}(2016)\citenamefont
  {Kharzeev}, \citenamefont {Liao}, \citenamefont {Voloshin},\ and\
  \citenamefont {Wang}}]{Kharzeev:2015znc}%
  \BibitemOpen
  \bibfield  {author} {\bibinfo {author} {\bibfnamefont {D.}~\bibnamefont
  {Kharzeev}}, \bibinfo {author} {\bibfnamefont {J.}~\bibnamefont {Liao}},
  \bibinfo {author} {\bibfnamefont {S.}~\bibnamefont {Voloshin}}, \ and\
  \bibinfo {author} {\bibfnamefont {G.}~\bibnamefont {Wang}},\ }\href {\doibase
  10.1016/j.ppnp.2016.01.001} {\bibfield  {journal} {\bibinfo  {journal} {Prog.
  Part. Nucl. Phys.}\ }\textbf {\bibinfo {volume} {88}},\ \bibinfo {pages} {1}
  (\bibinfo {year} {2016})},\ \Eprint {http://arxiv.org/abs/1511.04050}
  {arXiv:1511.04050 [hep-ph]} \BibitemShut {NoStop}%
\bibitem [{\citenamefont {Gusynin}\ \emph {et~al.}(1996)\citenamefont
  {Gusynin}, \citenamefont {Miransky},\ and\ \citenamefont
  {Shovkovy}}]{Gusynin:1995nb}%
  \BibitemOpen
  \bibfield  {author} {\bibinfo {author} {\bibfnamefont {V.}~\bibnamefont
  {Gusynin}}, \bibinfo {author} {\bibfnamefont {V.}~\bibnamefont {Miransky}}, \
  and\ \bibinfo {author} {\bibfnamefont {I.}~\bibnamefont {Shovkovy}},\ }\href
  {\doibase 10.1016/0550-3213(96)00021-1} {\bibfield  {journal} {\bibinfo
  {journal} {Nucl. Phys. B}\ }\textbf {\bibinfo {volume} {462}},\ \bibinfo
  {pages} {249} (\bibinfo {year} {1996})},\ \Eprint
  {http://arxiv.org/abs/hep-ph/9509320} {arXiv:hep-ph/9509320} \BibitemShut
  {NoStop}%
\bibitem [{\citenamefont {Sadofyev}\ and\ \citenamefont
  {Isachenkov}(2011)}]{Sadofyev:2010pr}%
  \BibitemOpen
  \bibfield  {author} {\bibinfo {author} {\bibfnamefont {A.}~\bibnamefont
  {Sadofyev}}\ and\ \bibinfo {author} {\bibfnamefont {M.}~\bibnamefont
  {Isachenkov}},\ }\href {\doibase 10.1016/j.physletb.2011.02.041} {\bibfield
  {journal} {\bibinfo  {journal} {Phys. Lett. B}\ }\textbf {\bibinfo {volume}
  {697}},\ \bibinfo {pages} {404} (\bibinfo {year} {2011})},\ \Eprint
  {http://arxiv.org/abs/1010.1550} {arXiv:1010.1550 [hep-th]} \BibitemShut
  {NoStop}%
\bibitem [{\citenamefont {Bali}\ \emph {et~al.}(2012)\citenamefont {Bali},
  \citenamefont {Bruckmann}, \citenamefont {Endrodi}, \citenamefont {Fodor},
  \citenamefont {Katz}, \citenamefont {Krieg}, \citenamefont {Schafer},\ and\
  \citenamefont {Szabo}}]{Bali:2011qj}%
  \BibitemOpen
  \bibfield  {author} {\bibinfo {author} {\bibfnamefont {G.}~\bibnamefont
  {Bali}}, \bibinfo {author} {\bibfnamefont {F.}~\bibnamefont {Bruckmann}},
  \bibinfo {author} {\bibfnamefont {G.}~\bibnamefont {Endrodi}}, \bibinfo
  {author} {\bibfnamefont {Z.}~\bibnamefont {Fodor}}, \bibinfo {author}
  {\bibfnamefont {S.}~\bibnamefont {Katz}}, \bibinfo {author} {\bibfnamefont
  {S.}~\bibnamefont {Krieg}}, \bibinfo {author} {\bibfnamefont
  {A.}~\bibnamefont {Schafer}}, \ and\ \bibinfo {author} {\bibfnamefont
  {K.}~\bibnamefont {Szabo}},\ }\href {\doibase 10.1007/JHEP02(2012)044}
  {\bibfield  {journal} {\bibinfo  {journal} {JHEP}\ }\textbf {\bibinfo
  {volume} {02}},\ \bibinfo {pages} {044} (\bibinfo {year} {2012})},\ \Eprint
  {http://arxiv.org/abs/1111.4956} {arXiv:1111.4956 [hep-lat]} \BibitemShut
  {NoStop}%
\bibitem [{\citenamefont {Kurian}\ \emph {et~al.}(2020)\citenamefont {Kurian},
  \citenamefont {Chandra},\ and\ \citenamefont {Das}}]{Kurian:2020kct}%
  \BibitemOpen
  \bibfield  {author} {\bibinfo {author} {\bibfnamefont {M.}~\bibnamefont
  {Kurian}}, \bibinfo {author} {\bibfnamefont {V.}~\bibnamefont {Chandra}}, \
  and\ \bibinfo {author} {\bibfnamefont {S.~K.}\ \bibnamefont {Das}},\ }\href
  {\doibase 10.1103/PhysRevD.101.094024} {\bibfield  {journal} {\bibinfo
  {journal} {Phys. Rev. D}\ }\textbf {\bibinfo {volume} {101}},\ \bibinfo
  {pages} {094024} (\bibinfo {year} {2020})},\ \Eprint
  {http://arxiv.org/abs/2002.03325} {arXiv:2002.03325 [nucl-th]} \BibitemShut
  {NoStop}%
\bibitem [{\citenamefont {Karmakar}\ \emph {et~al.}(2019)\citenamefont
  {Karmakar}, \citenamefont {Bandyopadhyay}, \citenamefont {Haque},\ and\
  \citenamefont {Mustafa}}]{Karmakar:2018aig}%
  \BibitemOpen
  \bibfield  {author} {\bibinfo {author} {\bibfnamefont {B.}~\bibnamefont
  {Karmakar}}, \bibinfo {author} {\bibfnamefont {A.}~\bibnamefont
  {Bandyopadhyay}}, \bibinfo {author} {\bibfnamefont {N.}~\bibnamefont
  {Haque}}, \ and\ \bibinfo {author} {\bibfnamefont {M.~G.}\ \bibnamefont
  {Mustafa}},\ }\href {\doibase 10.1140/epjc/s10052-019-7154-0} {\bibfield
  {journal} {\bibinfo  {journal} {Eur. Phys. J. C}\ }\textbf {\bibinfo {volume}
  {79}},\ \bibinfo {pages} {658} (\bibinfo {year} {2019})},\ \Eprint
  {http://arxiv.org/abs/1804.11336} {arXiv:1804.11336 [hep-ph]} \BibitemShut
  {NoStop}%
\bibitem [{\citenamefont {Ghosh}\ \emph {et~al.}(2020)\citenamefont {Ghosh},
  \citenamefont {Karmakar},\ and\ \citenamefont {Mustafa}}]{Ghosh:2019kmf}%
  \BibitemOpen
  \bibfield  {author} {\bibinfo {author} {\bibfnamefont {R.}~\bibnamefont
  {Ghosh}}, \bibinfo {author} {\bibfnamefont {B.}~\bibnamefont {Karmakar}}, \
  and\ \bibinfo {author} {\bibfnamefont {M.~G.}\ \bibnamefont {Mustafa}},\
  }\href {\doibase 10.1103/PhysRevD.101.056007} {\bibfield  {journal} {\bibinfo
   {journal} {Phys. Rev. D}\ }\textbf {\bibinfo {volume} {101}},\ \bibinfo
  {pages} {056007} (\bibinfo {year} {2020})},\ \Eprint
  {http://arxiv.org/abs/1911.00744} {arXiv:1911.00744 [hep-ph]} \BibitemShut
  {NoStop}%
\bibitem [{\citenamefont {Greif}\ \emph {et~al.}(2017)\citenamefont {Greif},
  \citenamefont {Greiner},\ and\ \citenamefont {Xu}}]{Greif:2017irh}%
  \BibitemOpen
  \bibfield  {author} {\bibinfo {author} {\bibfnamefont {M.}~\bibnamefont
  {Greif}}, \bibinfo {author} {\bibfnamefont {C.}~\bibnamefont {Greiner}}, \
  and\ \bibinfo {author} {\bibfnamefont {Z.}~\bibnamefont {Xu}},\ }\href
  {\doibase 10.1103/PhysRevC.96.014903} {\bibfield  {journal} {\bibinfo
  {journal} {Phys. Rev. C}\ }\textbf {\bibinfo {volume} {96}},\ \bibinfo
  {pages} {014903} (\bibinfo {year} {2017})},\ \Eprint
  {http://arxiv.org/abs/1704.06505} {arXiv:1704.06505 [hep-ph]} \BibitemShut
  {NoStop}%
\bibitem [{\citenamefont {Singh}\ \emph {et~al.}(2018)\citenamefont {Singh},
  \citenamefont {Thakur},\ and\ \citenamefont {Mishra}}]{Singh:2017nfa}%
  \BibitemOpen
  \bibfield  {author} {\bibinfo {author} {\bibfnamefont {B.}~\bibnamefont
  {Singh}}, \bibinfo {author} {\bibfnamefont {L.}~\bibnamefont {Thakur}}, \
  and\ \bibinfo {author} {\bibfnamefont {H.}~\bibnamefont {Mishra}},\ }\href
  {\doibase 10.1103/PhysRevD.97.096011} {\bibfield  {journal} {\bibinfo
  {journal} {Phys. Rev. D}\ }\textbf {\bibinfo {volume} {97}},\ \bibinfo
  {pages} {096011} (\bibinfo {year} {2018})},\ \Eprint
  {http://arxiv.org/abs/1711.03071} {arXiv:1711.03071 [hep-ph]} \BibitemShut
  {NoStop}%
\bibitem [{\citenamefont {Feng}\ and\ \citenamefont
  {Wang}(2017)}]{Feng:2017giy}%
  \BibitemOpen
  \bibfield  {author} {\bibinfo {author} {\bibfnamefont {B.}~\bibnamefont
  {Feng}}\ and\ \bibinfo {author} {\bibfnamefont {Z.}~\bibnamefont {Wang}},\
  }\href {\doibase 10.1103/PhysRevC.95.054912} {\bibfield  {journal} {\bibinfo
  {journal} {Phys. Rev. C}\ }\textbf {\bibinfo {volume} {95}},\ \bibinfo
  {pages} {054912} (\bibinfo {year} {2017})},\ \Eprint
  {http://arxiv.org/abs/1705.07842} {arXiv:1705.07842 [nucl-th]} \BibitemShut
  {NoStop}%
\bibitem [{\citenamefont {Rath}\ and\ \citenamefont
  {Patra}(2020)}]{Rath:2020idp}%
  \BibitemOpen
  \bibfield  {author} {\bibinfo {author} {\bibfnamefont {S.}~\bibnamefont
  {Rath}}\ and\ \bibinfo {author} {\bibfnamefont {B.~K.}\ \bibnamefont
  {Patra}},\ }\href {\doibase 10.1140/epjc/s10052-020-8331-x} {\bibfield
  {journal} {\bibinfo  {journal} {Eur. Phys. J. C}\ }\textbf {\bibinfo {volume}
  {80}},\ \bibinfo {pages} {747} (\bibinfo {year} {2020})},\ \Eprint
  {http://arxiv.org/abs/2005.00997} {arXiv:2005.00997 [hep-ph]} \BibitemShut
  {NoStop}%
\bibitem [{\citenamefont {Tuchin}(2013)}]{Tuchin:2013apa}%
  \BibitemOpen
  \bibfield  {author} {\bibinfo {author} {\bibfnamefont {K.}~\bibnamefont
  {Tuchin}},\ }\href {\doibase 10.1103/PhysRevC.88.024911} {\bibfield
  {journal} {\bibinfo  {journal} {Phys. Rev. C}\ }\textbf {\bibinfo {volume}
  {88}},\ \bibinfo {pages} {024911} (\bibinfo {year} {2013})},\ \Eprint
  {http://arxiv.org/abs/1305.5806} {arXiv:1305.5806 [hep-ph]} \BibitemShut
  {NoStop}%
\bibitem [{\citenamefont {Tuchin}(2016)}]{Tuchin:2015oka}%
  \BibitemOpen
  \bibfield  {author} {\bibinfo {author} {\bibfnamefont {K.}~\bibnamefont
  {Tuchin}},\ }\href {\doibase 10.1103/PhysRevC.93.014905} {\bibfield
  {journal} {\bibinfo  {journal} {Phys. Rev. C}\ }\textbf {\bibinfo {volume}
  {93}},\ \bibinfo {pages} {014905} (\bibinfo {year} {2016})},\ \Eprint
  {http://arxiv.org/abs/1508.06925} {arXiv:1508.06925 [hep-ph]} \BibitemShut
  {NoStop}%
\bibitem [{\citenamefont {McLerran}\ and\ \citenamefont
  {Skokov}(2014)}]{McLerran:2013hla}%
  \BibitemOpen
  \bibfield  {author} {\bibinfo {author} {\bibfnamefont {L.}~\bibnamefont
  {McLerran}}\ and\ \bibinfo {author} {\bibfnamefont {V.}~\bibnamefont
  {Skokov}},\ }\href {\doibase 10.1016/j.nuclphysa.2014.05.008} {\bibfield
  {journal} {\bibinfo  {journal} {Nucl. Phys. A}\ }\textbf {\bibinfo {volume}
  {929}},\ \bibinfo {pages} {184} (\bibinfo {year} {2014})},\ \Eprint
  {http://arxiv.org/abs/1305.0774} {arXiv:1305.0774 [hep-ph]} \BibitemShut
  {NoStop}%
\bibitem [{\citenamefont {Zakharov}(2014)}]{Zakharov:2014dia}%
  \BibitemOpen
  \bibfield  {author} {\bibinfo {author} {\bibfnamefont {B.}~\bibnamefont
  {Zakharov}},\ }\href {\doibase 10.1016/j.physletb.2014.08.068} {\bibfield
  {journal} {\bibinfo  {journal} {Phys. Lett. B}\ }\textbf {\bibinfo {volume}
  {737}},\ \bibinfo {pages} {262} (\bibinfo {year} {2014})},\ \Eprint
  {http://arxiv.org/abs/1404.5047} {arXiv:1404.5047 [hep-ph]} \BibitemShut
  {NoStop}%
\bibitem [{\citenamefont {Gursoy}\ \emph {et~al.}(2014)\citenamefont {Gursoy},
  \citenamefont {Kharzeev},\ and\ \citenamefont {Rajagopal}}]{Gursoy:2014aka}%
  \BibitemOpen
  \bibfield  {author} {\bibinfo {author} {\bibfnamefont {U.}~\bibnamefont
  {Gursoy}}, \bibinfo {author} {\bibfnamefont {D.}~\bibnamefont {Kharzeev}}, \
  and\ \bibinfo {author} {\bibfnamefont {K.}~\bibnamefont {Rajagopal}},\ }\href
  {\doibase 10.1103/PhysRevC.89.054905} {\bibfield  {journal} {\bibinfo
  {journal} {Phys. Rev. C}\ }\textbf {\bibinfo {volume} {89}},\ \bibinfo
  {pages} {054905} (\bibinfo {year} {2014})},\ \Eprint
  {http://arxiv.org/abs/1401.3805} {arXiv:1401.3805 [hep-ph]} \BibitemShut
  {NoStop}%
\bibitem [{\citenamefont {Jiang}\ \emph {et~al.}(2020)\citenamefont {Jiang},
  \citenamefont {Shi}, \citenamefont {Hou},\ and\ \citenamefont
  {Li}}]{Jiang:2020lgw}%
  \BibitemOpen
  \bibfield  {author} {\bibinfo {author} {\bibfnamefont {B.-f.}\ \bibnamefont
  {Jiang}}, \bibinfo {author} {\bibfnamefont {S.-w.}\ \bibnamefont {Shi}},
  \bibinfo {author} {\bibfnamefont {D.-f.}\ \bibnamefont {Hou}}, \ and\
  \bibinfo {author} {\bibfnamefont {J.-r.}\ \bibnamefont {Li}},\ }\href@noop {}
  {\  (\bibinfo {year} {2020})},\ \Eprint {http://arxiv.org/abs/2007.00902}
  {arXiv:2007.00902 [hep-ph]} \BibitemShut {NoStop}%
\bibitem [{\citenamefont {Greif}\ \emph {et~al.}(2014)\citenamefont {Greif},
  \citenamefont {Bouras}, \citenamefont {Greiner},\ and\ \citenamefont
  {Xu}}]{Greif:2014oia}%
  \BibitemOpen
  \bibfield  {author} {\bibinfo {author} {\bibfnamefont {M.}~\bibnamefont
  {Greif}}, \bibinfo {author} {\bibfnamefont {I.}~\bibnamefont {Bouras}},
  \bibinfo {author} {\bibfnamefont {C.}~\bibnamefont {Greiner}}, \ and\
  \bibinfo {author} {\bibfnamefont {Z.}~\bibnamefont {Xu}},\ }\href {\doibase
  10.1103/PhysRevD.90.094014} {\bibfield  {journal} {\bibinfo  {journal} {Phys.
  Rev. D}\ }\textbf {\bibinfo {volume} {90}},\ \bibinfo {pages} {094014}
  (\bibinfo {year} {2014})},\ \Eprint {http://arxiv.org/abs/1408.7049}
  {arXiv:1408.7049 [nucl-th]} \BibitemShut {NoStop}%
\bibitem [{\citenamefont {Cassing}\ \emph {et~al.}(2013)\citenamefont
  {Cassing}, \citenamefont {Linnyk}, \citenamefont {Steinert},\ and\
  \citenamefont {Ozvenchuk}}]{Cassing:2013iz}%
  \BibitemOpen
  \bibfield  {author} {\bibinfo {author} {\bibfnamefont {W.}~\bibnamefont
  {Cassing}}, \bibinfo {author} {\bibfnamefont {O.}~\bibnamefont {Linnyk}},
  \bibinfo {author} {\bibfnamefont {T.}~\bibnamefont {Steinert}}, \ and\
  \bibinfo {author} {\bibfnamefont {V.}~\bibnamefont {Ozvenchuk}},\ }\href
  {\doibase 10.1103/PhysRevLett.110.182301} {\bibfield  {journal} {\bibinfo
  {journal} {Phys. Rev. Lett.}\ }\textbf {\bibinfo {volume} {110}},\ \bibinfo
  {pages} {182301} (\bibinfo {year} {2013})},\ \Eprint
  {http://arxiv.org/abs/1302.0906} {arXiv:1302.0906 [hep-ph]} \BibitemShut
  {NoStop}%
\bibitem [{\citenamefont {Puglisi}\ \emph {et~al.}(2014)\citenamefont
  {Puglisi}, \citenamefont {Plumari},\ and\ \citenamefont
  {Greco}}]{Puglisi:2014sha}%
  \BibitemOpen
  \bibfield  {author} {\bibinfo {author} {\bibfnamefont {A.}~\bibnamefont
  {Puglisi}}, \bibinfo {author} {\bibfnamefont {S.}~\bibnamefont {Plumari}}, \
  and\ \bibinfo {author} {\bibfnamefont {V.}~\bibnamefont {Greco}},\ }\href
  {\doibase 10.1103/PhysRevD.90.114009} {\bibfield  {journal} {\bibinfo
  {journal} {Phys. Rev. D}\ }\textbf {\bibinfo {volume} {90}},\ \bibinfo
  {pages} {114009} (\bibinfo {year} {2014})},\ \Eprint
  {http://arxiv.org/abs/1408.7043} {arXiv:1408.7043 [hep-ph]} \BibitemShut
  {NoStop}%
\bibitem [{\citenamefont {Mitra}\ and\ \citenamefont
  {Chandra}(2016)}]{Mitra:2016zdw}%
  \BibitemOpen
  \bibfield  {author} {\bibinfo {author} {\bibfnamefont {S.}~\bibnamefont
  {Mitra}}\ and\ \bibinfo {author} {\bibfnamefont {V.}~\bibnamefont
  {Chandra}},\ }\href {\doibase 10.1103/PhysRevD.94.034025} {\bibfield
  {journal} {\bibinfo  {journal} {Phys. Rev. D}\ }\textbf {\bibinfo {volume}
  {94}},\ \bibinfo {pages} {034025} (\bibinfo {year} {2016})},\ \Eprint
  {http://arxiv.org/abs/1606.08556} {arXiv:1606.08556 [nucl-th]} \BibitemShut
  {NoStop}%
\bibitem [{\citenamefont {Thakur}\ \emph {et~al.}(2017)\citenamefont {Thakur},
  \citenamefont {Srivastava}, \citenamefont {Kadam}, \citenamefont {George},\
  and\ \citenamefont {Mishra}}]{Thakur:2017hfc}%
  \BibitemOpen
  \bibfield  {author} {\bibinfo {author} {\bibfnamefont {L.}~\bibnamefont
  {Thakur}}, \bibinfo {author} {\bibfnamefont {P.}~\bibnamefont {Srivastava}},
  \bibinfo {author} {\bibfnamefont {G.~P.}\ \bibnamefont {Kadam}}, \bibinfo
  {author} {\bibfnamefont {M.}~\bibnamefont {George}}, \ and\ \bibinfo {author}
  {\bibfnamefont {H.}~\bibnamefont {Mishra}},\ }\href {\doibase
  10.1103/PhysRevD.95.096009} {\bibfield  {journal} {\bibinfo  {journal} {Phys.
  Rev. D}\ }\textbf {\bibinfo {volume} {95}},\ \bibinfo {pages} {096009}
  (\bibinfo {year} {2017})},\ \Eprint {http://arxiv.org/abs/1703.03142}
  {arXiv:1703.03142 [hep-ph]} \BibitemShut {NoStop}%
\bibitem [{\citenamefont {Astrakhantsev}\ \emph {et~al.}(2020)\citenamefont
  {Astrakhantsev}, \citenamefont {Braguta}, \citenamefont {D'Elia},
  \citenamefont {Kotov}, \citenamefont {Nikolaev},\ and\ \citenamefont
  {Sanfilippo}}]{Astrakhantsev:2019zkr}%
  \BibitemOpen
  \bibfield  {author} {\bibinfo {author} {\bibfnamefont {N.}~\bibnamefont
  {Astrakhantsev}}, \bibinfo {author} {\bibfnamefont {V.~V.}\ \bibnamefont
  {Braguta}}, \bibinfo {author} {\bibfnamefont {M.}~\bibnamefont {D'Elia}},
  \bibinfo {author} {\bibfnamefont {A.~Y.}\ \bibnamefont {Kotov}}, \bibinfo
  {author} {\bibfnamefont {A.~A.}\ \bibnamefont {Nikolaev}}, \ and\ \bibinfo
  {author} {\bibfnamefont {F.}~\bibnamefont {Sanfilippo}},\ }\href {\doibase
  10.1103/PhysRevD.102.054516} {\bibfield  {journal} {\bibinfo  {journal}
  {Phys. Rev. D}\ }\textbf {\bibinfo {volume} {102}},\ \bibinfo {pages}
  {054516} (\bibinfo {year} {2020})},\ \Eprint
  {http://arxiv.org/abs/1910.08516} {arXiv:1910.08516 [hep-lat]} \BibitemShut
  {NoStop}%
\bibitem [{\citenamefont {Amato}\ \emph {et~al.}(2013)\citenamefont {Amato},
  \citenamefont {Aarts}, \citenamefont {Allton}, \citenamefont {Giudice},
  \citenamefont {Hands},\ and\ \citenamefont {Skullerud}}]{Amato:2013naa}%
  \BibitemOpen
  \bibfield  {author} {\bibinfo {author} {\bibfnamefont {A.}~\bibnamefont
  {Amato}}, \bibinfo {author} {\bibfnamefont {G.}~\bibnamefont {Aarts}},
  \bibinfo {author} {\bibfnamefont {C.}~\bibnamefont {Allton}}, \bibinfo
  {author} {\bibfnamefont {P.}~\bibnamefont {Giudice}}, \bibinfo {author}
  {\bibfnamefont {S.}~\bibnamefont {Hands}}, \ and\ \bibinfo {author}
  {\bibfnamefont {J.-I.}\ \bibnamefont {Skullerud}},\ }\href {\doibase
  10.1103/PhysRevLett.111.172001} {\bibfield  {journal} {\bibinfo  {journal}
  {Phys. Rev. Lett.}\ }\textbf {\bibinfo {volume} {111}},\ \bibinfo {pages}
  {172001} (\bibinfo {year} {2013})},\ \Eprint {http://arxiv.org/abs/1307.6763}
  {arXiv:1307.6763 [hep-lat]} \BibitemShut {NoStop}%
\bibitem [{\citenamefont {Francis}\ and\ \citenamefont
  {Kaczmarek}(2012)}]{Francis:2011bt}%
  \BibitemOpen
  \bibfield  {author} {\bibinfo {author} {\bibfnamefont {A.}~\bibnamefont
  {Francis}}\ and\ \bibinfo {author} {\bibfnamefont {O.}~\bibnamefont
  {Kaczmarek}},\ }\href {\doibase 10.1016/j.ppnp.2011.12.020} {\bibfield
  {journal} {\bibinfo  {journal} {Prog. Part. Nucl. Phys.}\ }\textbf {\bibinfo
  {volume} {67}},\ \bibinfo {pages} {212} (\bibinfo {year} {2012})},\ \Eprint
  {http://arxiv.org/abs/1112.4802} {arXiv:1112.4802 [hep-lat]} \BibitemShut
  {NoStop}%
\bibitem [{\citenamefont {Kubo}(1957)}]{Kubo:1957mj}%
  \BibitemOpen
  \bibfield  {author} {\bibinfo {author} {\bibfnamefont {R.}~\bibnamefont
  {Kubo}},\ }\href {\doibase 10.1143/JPSJ.12.570} {\bibfield  {journal}
  {\bibinfo  {journal} {J. Phys. Soc. Jap.}\ }\textbf {\bibinfo {volume}
  {12}},\ \bibinfo {pages} {570} (\bibinfo {year} {1957})}\BibitemShut
  {NoStop}%
\bibitem [{\citenamefont {Finazzo}\ and\ \citenamefont
  {Rougemont}(2016)}]{Finazzo:2015xwa}%
  \BibitemOpen
  \bibfield  {author} {\bibinfo {author} {\bibfnamefont {S.~I.}\ \bibnamefont
  {Finazzo}}\ and\ \bibinfo {author} {\bibfnamefont {R.}~\bibnamefont
  {Rougemont}},\ }\href {\doibase 10.1103/PhysRevD.93.034017} {\bibfield
  {journal} {\bibinfo  {journal} {Phys. Rev. D}\ }\textbf {\bibinfo {volume}
  {93}},\ \bibinfo {pages} {034017} (\bibinfo {year} {2016})},\ \Eprint
  {http://arxiv.org/abs/1510.03321} {arXiv:1510.03321 [hep-ph]} \BibitemShut
  {NoStop}%
\bibitem [{\citenamefont {Jain}(2010)}]{Jain:2010ip}%
  \BibitemOpen
  \bibfield  {author} {\bibinfo {author} {\bibfnamefont {S.}~\bibnamefont
  {Jain}},\ }\href {\doibase 10.1007/JHEP11(2010)092} {\bibfield  {journal}
  {\bibinfo  {journal} {JHEP}\ }\textbf {\bibinfo {volume} {11}},\ \bibinfo
  {pages} {092} (\bibinfo {year} {2010})},\ \Eprint
  {http://arxiv.org/abs/1008.2944} {arXiv:1008.2944 [hep-th]} \BibitemShut
  {NoStop}%
\bibitem [{\citenamefont {Yin}(2014)}]{Yin:2013kya}%
  \BibitemOpen
  \bibfield  {author} {\bibinfo {author} {\bibfnamefont {Y.}~\bibnamefont
  {Yin}},\ }\href {\doibase 10.1103/PhysRevC.90.044903} {\bibfield  {journal}
  {\bibinfo  {journal} {Phys. Rev. C}\ }\textbf {\bibinfo {volume} {90}},\
  \bibinfo {pages} {044903} (\bibinfo {year} {2014})},\ \Eprint
  {http://arxiv.org/abs/1312.4434} {arXiv:1312.4434 [nucl-th]} \BibitemShut
  {NoStop}%
\bibitem [{\citenamefont {Hirono}\ \emph {et~al.}(2014)\citenamefont {Hirono},
  \citenamefont {Hongo},\ and\ \citenamefont {Hirano}}]{Hirono:2012rt}%
  \BibitemOpen
  \bibfield  {author} {\bibinfo {author} {\bibfnamefont {Y.}~\bibnamefont
  {Hirono}}, \bibinfo {author} {\bibfnamefont {M.}~\bibnamefont {Hongo}}, \
  and\ \bibinfo {author} {\bibfnamefont {T.}~\bibnamefont {Hirano}},\ }\href
  {\doibase 10.1103/PhysRevC.90.021903} {\bibfield  {journal} {\bibinfo
  {journal} {Phys. Rev. C}\ }\textbf {\bibinfo {volume} {90}},\ \bibinfo
  {pages} {021903} (\bibinfo {year} {2014})},\ \Eprint
  {http://arxiv.org/abs/1211.1114} {arXiv:1211.1114 [nucl-th]} \BibitemShut
  {NoStop}%
\bibitem [{\citenamefont {Hattori}\ and\ \citenamefont
  {Satow}(2016)}]{Hattori:2016cnt}%
  \BibitemOpen
  \bibfield  {author} {\bibinfo {author} {\bibfnamefont {K.}~\bibnamefont
  {Hattori}}\ and\ \bibinfo {author} {\bibfnamefont {D.}~\bibnamefont
  {Satow}},\ }\href {\doibase 10.1103/PhysRevD.94.114032} {\bibfield  {journal}
  {\bibinfo  {journal} {Phys. Rev. D}\ }\textbf {\bibinfo {volume} {94}},\
  \bibinfo {pages} {114032} (\bibinfo {year} {2016})},\ \Eprint
  {http://arxiv.org/abs/1610.06818} {arXiv:1610.06818 [hep-ph]} \BibitemShut
  {NoStop}%
\bibitem [{\citenamefont {Hattori}\ \emph
  {et~al.}(2017{\natexlab{a}})\citenamefont {Hattori}, \citenamefont {Li},
  \citenamefont {Satow},\ and\ \citenamefont {Yee}}]{Hattori:2016lqx}%
  \BibitemOpen
  \bibfield  {author} {\bibinfo {author} {\bibfnamefont {K.}~\bibnamefont
  {Hattori}}, \bibinfo {author} {\bibfnamefont {S.}~\bibnamefont {Li}},
  \bibinfo {author} {\bibfnamefont {D.}~\bibnamefont {Satow}}, \ and\ \bibinfo
  {author} {\bibfnamefont {H.-U.}\ \bibnamefont {Yee}},\ }\href {\doibase
  10.1103/PhysRevD.95.076008} {\bibfield  {journal} {\bibinfo  {journal} {Phys.
  Rev. D}\ }\textbf {\bibinfo {volume} {95}},\ \bibinfo {pages} {076008}
  (\bibinfo {year} {2017}{\natexlab{a}})},\ \Eprint
  {http://arxiv.org/abs/1610.06839} {arXiv:1610.06839 [hep-ph]} \BibitemShut
  {NoStop}%
\bibitem [{\citenamefont {Fukushima}\ and\ \citenamefont
  {Hidaka}(2018)}]{Fukushima:2017lvb}%
  \BibitemOpen
  \bibfield  {author} {\bibinfo {author} {\bibfnamefont {K.}~\bibnamefont
  {Fukushima}}\ and\ \bibinfo {author} {\bibfnamefont {Y.}~\bibnamefont
  {Hidaka}},\ }\href {\doibase 10.1103/PhysRevLett.120.162301} {\bibfield
  {journal} {\bibinfo  {journal} {Phys. Rev. Lett.}\ }\textbf {\bibinfo
  {volume} {120}},\ \bibinfo {pages} {162301} (\bibinfo {year} {2018})},\
  \Eprint {http://arxiv.org/abs/1711.01472} {arXiv:1711.01472 [hep-ph]}
  \BibitemShut {NoStop}%
\bibitem [{\citenamefont {Kurian}\ and\ \citenamefont
  {Chandra}(2017)}]{Kurian:2017yxj}%
  \BibitemOpen
  \bibfield  {author} {\bibinfo {author} {\bibfnamefont {M.}~\bibnamefont
  {Kurian}}\ and\ \bibinfo {author} {\bibfnamefont {V.}~\bibnamefont
  {Chandra}},\ }\href {\doibase 10.1103/PhysRevD.96.114026} {\bibfield
  {journal} {\bibinfo  {journal} {Phys. Rev. D}\ }\textbf {\bibinfo {volume}
  {96}},\ \bibinfo {pages} {114026} (\bibinfo {year} {2017})},\ \Eprint
  {http://arxiv.org/abs/1709.08320} {arXiv:1709.08320 [nucl-th]} \BibitemShut
  {NoStop}%
\bibitem [{\citenamefont {Ghosh}\ \emph {et~al.}(2019)\citenamefont {Ghosh},
  \citenamefont {Bandyopadhyay}, \citenamefont {Farias}, \citenamefont {Dey},\
  and\ \citenamefont {Krein}}]{Ghosh:2019ubc}%
  \BibitemOpen
  \bibfield  {author} {\bibinfo {author} {\bibfnamefont {S.}~\bibnamefont
  {Ghosh}}, \bibinfo {author} {\bibfnamefont {A.}~\bibnamefont
  {Bandyopadhyay}}, \bibinfo {author} {\bibfnamefont {R.~L.}\ \bibnamefont
  {Farias}}, \bibinfo {author} {\bibfnamefont {J.}~\bibnamefont {Dey}}, \ and\
  \bibinfo {author} {\bibfnamefont {G.~a.}\ \bibnamefont {Krein}},\ }\href@noop
  {} {\  (\bibinfo {year} {2019})},\ \Eprint {http://arxiv.org/abs/1911.10005}
  {arXiv:1911.10005 [hep-ph]} \BibitemShut {NoStop}%
\bibitem [{\citenamefont {Feng}(2017)}]{Feng:2017tsh}%
  \BibitemOpen
  \bibfield  {author} {\bibinfo {author} {\bibfnamefont {B.}~\bibnamefont
  {Feng}},\ }\href {\doibase 10.1103/PhysRevD.96.036009} {\bibfield  {journal}
  {\bibinfo  {journal} {Phys. Rev. D}\ }\textbf {\bibinfo {volume} {96}},\
  \bibinfo {pages} {036009} (\bibinfo {year} {2017})}\BibitemShut {NoStop}%
\bibitem [{\citenamefont {Das}\ \emph {et~al.}(2020)\citenamefont {Das},
  \citenamefont {Mishra},\ and\ \citenamefont {Mohapatra}}]{Das:2019ppb}%
  \BibitemOpen
  \bibfield  {author} {\bibinfo {author} {\bibfnamefont {A.}~\bibnamefont
  {Das}}, \bibinfo {author} {\bibfnamefont {H.}~\bibnamefont {Mishra}}, \ and\
  \bibinfo {author} {\bibfnamefont {R.~K.}\ \bibnamefont {Mohapatra}},\ }\href
  {\doibase 10.1103/PhysRevD.101.034027} {\bibfield  {journal} {\bibinfo
  {journal} {Phys. Rev. D}\ }\textbf {\bibinfo {volume} {101}},\ \bibinfo
  {pages} {034027} (\bibinfo {year} {2020})},\ \Eprint
  {http://arxiv.org/abs/1907.05298} {arXiv:1907.05298 [hep-ph]} \BibitemShut
  {NoStop}%
\bibitem [{\citenamefont {Das}\ \emph {et~al.}(2019)\citenamefont {Das},
  \citenamefont {Mishra},\ and\ \citenamefont {Mohapatra}}]{Das:2019wjg}%
  \BibitemOpen
  \bibfield  {author} {\bibinfo {author} {\bibfnamefont {A.}~\bibnamefont
  {Das}}, \bibinfo {author} {\bibfnamefont {H.}~\bibnamefont {Mishra}}, \ and\
  \bibinfo {author} {\bibfnamefont {R.~K.}\ \bibnamefont {Mohapatra}},\ }\href
  {\doibase 10.1103/PhysRevD.99.094031} {\bibfield  {journal} {\bibinfo
  {journal} {Phys. Rev. D}\ }\textbf {\bibinfo {volume} {99}},\ \bibinfo
  {pages} {094031} (\bibinfo {year} {2019})},\ \Eprint
  {http://arxiv.org/abs/1903.03938} {arXiv:1903.03938 [hep-ph]} \BibitemShut
  {NoStop}%
\bibitem [{\citenamefont {Thakur}\ and\ \citenamefont
  {Srivastava}(2019)}]{Thakur:2019bnf}%
  \BibitemOpen
  \bibfield  {author} {\bibinfo {author} {\bibfnamefont {L.}~\bibnamefont
  {Thakur}}\ and\ \bibinfo {author} {\bibfnamefont {P.}~\bibnamefont
  {Srivastava}},\ }\href {\doibase 10.1103/PhysRevD.100.076016} {\bibfield
  {journal} {\bibinfo  {journal} {Phys. Rev. D}\ }\textbf {\bibinfo {volume}
  {100}},\ \bibinfo {pages} {076016} (\bibinfo {year} {2019})},\ \Eprint
  {http://arxiv.org/abs/1910.12087} {arXiv:1910.12087 [hep-ph]} \BibitemShut
  {NoStop}%
\bibitem [{\citenamefont {Chatterjee}\ \emph {et~al.}(2019)\citenamefont
  {Chatterjee}, \citenamefont {Rath}, \citenamefont {Sarwar},\ and\
  \citenamefont {Sahoo}}]{Chatterjee:2019nld}%
  \BibitemOpen
  \bibfield  {author} {\bibinfo {author} {\bibfnamefont {B.}~\bibnamefont
  {Chatterjee}}, \bibinfo {author} {\bibfnamefont {R.}~\bibnamefont {Rath}},
  \bibinfo {author} {\bibfnamefont {G.}~\bibnamefont {Sarwar}}, \ and\ \bibinfo
  {author} {\bibfnamefont {R.}~\bibnamefont {Sahoo}},\ }\href@noop {} {\
  (\bibinfo {year} {2019})},\ \Eprint {http://arxiv.org/abs/1908.01121}
  {arXiv:1908.01121 [hep-ph]} \BibitemShut {NoStop}%
\bibitem [{\citenamefont {Dey}\ \emph {et~al.}(2019)\citenamefont {Dey},
  \citenamefont {Satapathy}, \citenamefont {Murmu},\ and\ \citenamefont
  {Ghosh}}]{Dey:2019axu}%
  \BibitemOpen
  \bibfield  {author} {\bibinfo {author} {\bibfnamefont {J.}~\bibnamefont
  {Dey}}, \bibinfo {author} {\bibfnamefont {S.}~\bibnamefont {Satapathy}},
  \bibinfo {author} {\bibfnamefont {P.}~\bibnamefont {Murmu}}, \ and\ \bibinfo
  {author} {\bibfnamefont {S.}~\bibnamefont {Ghosh}},\ }\href@noop {} {\
  (\bibinfo {year} {2019})},\ \Eprint {http://arxiv.org/abs/1907.11164}
  {arXiv:1907.11164 [hep-ph]} \BibitemShut {NoStop}%
\bibitem [{\citenamefont {Kurian}(2020)}]{Kurian:2020qjr}%
  \BibitemOpen
  \bibfield  {author} {\bibinfo {author} {\bibfnamefont {M.}~\bibnamefont
  {Kurian}},\ }\href {\doibase 10.1103/PhysRevD.102.014041} {\bibfield
  {journal} {\bibinfo  {journal} {Phys. Rev. D}\ }\textbf {\bibinfo {volume}
  {102}},\ \bibinfo {pages} {014041} (\bibinfo {year} {2020})},\ \Eprint
  {http://arxiv.org/abs/2005.04247} {arXiv:2005.04247 [nucl-th]} \BibitemShut
  {NoStop}%
\bibitem [{\citenamefont {Kalikotay}\ \emph {et~al.}(2020)\citenamefont
  {Kalikotay}, \citenamefont {Ghosh}, \citenamefont {Chaudhuri}, \citenamefont
  {Roy},\ and\ \citenamefont {Sarkar}}]{Kalikotay:2020snc}%
  \BibitemOpen
  \bibfield  {author} {\bibinfo {author} {\bibfnamefont {P.}~\bibnamefont
  {Kalikotay}}, \bibinfo {author} {\bibfnamefont {S.}~\bibnamefont {Ghosh}},
  \bibinfo {author} {\bibfnamefont {N.}~\bibnamefont {Chaudhuri}}, \bibinfo
  {author} {\bibfnamefont {P.}~\bibnamefont {Roy}}, \ and\ \bibinfo {author}
  {\bibfnamefont {S.}~\bibnamefont {Sarkar}},\ }\href {\doibase
  10.1103/PhysRevD.102.076007} {\bibfield  {journal} {\bibinfo  {journal}
  {Phys. Rev. D}\ }\textbf {\bibinfo {volume} {102}},\ \bibinfo {pages}
  {076007} (\bibinfo {year} {2020})},\ \Eprint
  {http://arxiv.org/abs/2009.10493} {arXiv:2009.10493 [hep-ph]} \BibitemShut
  {NoStop}%
\bibitem [{\citenamefont {Hongo}\ \emph {et~al.}(2017)\citenamefont {Hongo},
  \citenamefont {Hirono},\ and\ \citenamefont {Hirano}}]{Hongo:2013cqa}%
  \BibitemOpen
  \bibfield  {author} {\bibinfo {author} {\bibfnamefont {M.}~\bibnamefont
  {Hongo}}, \bibinfo {author} {\bibfnamefont {Y.}~\bibnamefont {Hirono}}, \
  and\ \bibinfo {author} {\bibfnamefont {T.}~\bibnamefont {Hirano}},\ }\href
  {\doibase 10.1016/j.physletb.2017.10.028} {\bibfield  {journal} {\bibinfo
  {journal} {Phys. Lett. B}\ }\textbf {\bibinfo {volume} {775}},\ \bibinfo
  {pages} {266} (\bibinfo {year} {2017})},\ \Eprint
  {http://arxiv.org/abs/1309.2823} {arXiv:1309.2823 [nucl-th]} \BibitemShut
  {NoStop}%
\bibitem [{\citenamefont {Anderson}\ and\ \citenamefont
  {Witting}(1974)}]{anderson1974relativistic}%
  \BibitemOpen
  \bibfield  {author} {\bibinfo {author} {\bibfnamefont {J.~L.}\ \bibnamefont
  {Anderson}}\ and\ \bibinfo {author} {\bibfnamefont {H.}~\bibnamefont
  {Witting}},\ }\href@noop {} {\bibfield  {journal} {\bibinfo  {journal}
  {Physica}\ }\textbf {\bibinfo {volume} {74}},\ \bibinfo {pages} {466}
  (\bibinfo {year} {1974})}\BibitemShut {NoStop}%
\bibitem [{\citenamefont {Bhatnagar}\ \emph {et~al.}(1954)\citenamefont
  {Bhatnagar}, \citenamefont {Gross},\ and\ \citenamefont
  {Krook}}]{Bhatnagar:1954zz}%
  \BibitemOpen
  \bibfield  {author} {\bibinfo {author} {\bibfnamefont {P.}~\bibnamefont
  {Bhatnagar}}, \bibinfo {author} {\bibfnamefont {E.}~\bibnamefont {Gross}}, \
  and\ \bibinfo {author} {\bibfnamefont {M.}~\bibnamefont {Krook}},\ }\href
  {\doibase 10.1103/PhysRev.94.511} {\bibfield  {journal} {\bibinfo  {journal}
  {Phys. Rev.}\ }\textbf {\bibinfo {volume} {94}},\ \bibinfo {pages} {511}
  (\bibinfo {year} {1954})}\BibitemShut {NoStop}%
\bibitem [{\citenamefont {Dash}\ \emph {et~al.}(2020)\citenamefont {Dash},
  \citenamefont {Samanta}, \citenamefont {Dey}, \citenamefont {Gangopadhyaya},
  \citenamefont {Ghosh},\ and\ \citenamefont {Roy}}]{Dash:2020vxk}%
  \BibitemOpen
  \bibfield  {author} {\bibinfo {author} {\bibfnamefont {A.}~\bibnamefont
  {Dash}}, \bibinfo {author} {\bibfnamefont {S.}~\bibnamefont {Samanta}},
  \bibinfo {author} {\bibfnamefont {J.}~\bibnamefont {Dey}}, \bibinfo {author}
  {\bibfnamefont {U.}~\bibnamefont {Gangopadhyaya}}, \bibinfo {author}
  {\bibfnamefont {S.}~\bibnamefont {Ghosh}}, \ and\ \bibinfo {author}
  {\bibfnamefont {V.}~\bibnamefont {Roy}},\ }\href {\doibase
  10.1103/PhysRevD.102.016016} {\bibfield  {journal} {\bibinfo  {journal}
  {Phys. Rev. D}\ }\textbf {\bibinfo {volume} {102}},\ \bibinfo {pages}
  {016016} (\bibinfo {year} {2020})},\ \Eprint
  {http://arxiv.org/abs/2002.08781} {arXiv:2002.08781 [nucl-th]} \BibitemShut
  {NoStop}%
\bibitem [{\citenamefont {Ghosh}\ and\ \citenamefont
  {Ghosh}(2020)}]{Ghosh:2020wqx}%
  \BibitemOpen
  \bibfield  {author} {\bibinfo {author} {\bibfnamefont {S.}~\bibnamefont
  {Ghosh}}\ and\ \bibinfo {author} {\bibfnamefont {S.}~\bibnamefont {Ghosh}},\
  }\href@noop {} {\  (\bibinfo {year} {2020})},\ \Eprint
  {http://arxiv.org/abs/2011.04261} {arXiv:2011.04261 [hep-ph]} \BibitemShut
  {NoStop}%
\bibitem [{\citenamefont {Panda}\ \emph {et~al.}(2020)\citenamefont {Panda},
  \citenamefont {Dash}, \citenamefont {Biswas},\ and\ \citenamefont
  {Roy}}]{Panda:2020zhr}%
  \BibitemOpen
  \bibfield  {author} {\bibinfo {author} {\bibfnamefont {A.~K.}\ \bibnamefont
  {Panda}}, \bibinfo {author} {\bibfnamefont {A.}~\bibnamefont {Dash}},
  \bibinfo {author} {\bibfnamefont {R.}~\bibnamefont {Biswas}}, \ and\ \bibinfo
  {author} {\bibfnamefont {V.}~\bibnamefont {Roy}},\ }\href@noop {} {\
  (\bibinfo {year} {2020})},\ \Eprint {http://arxiv.org/abs/2011.01606}
  {arXiv:2011.01606 [nucl-th]} \BibitemShut {NoStop}%
\bibitem [{\citenamefont {Huang}\ \emph {et~al.}(2011)\citenamefont {Huang},
  \citenamefont {Sedrakian},\ and\ \citenamefont {Rischke}}]{Huang:2011dc}%
  \BibitemOpen
  \bibfield  {author} {\bibinfo {author} {\bibfnamefont {X.-G.}\ \bibnamefont
  {Huang}}, \bibinfo {author} {\bibfnamefont {A.}~\bibnamefont {Sedrakian}}, \
  and\ \bibinfo {author} {\bibfnamefont {D.~H.}\ \bibnamefont {Rischke}},\
  }\href {\doibase 10.1016/j.aop.2011.08.001} {\bibfield  {journal} {\bibinfo
  {journal} {Annals Phys.}\ }\textbf {\bibinfo {volume} {326}},\ \bibinfo
  {pages} {3075} (\bibinfo {year} {2011})},\ \Eprint
  {http://arxiv.org/abs/1108.0602} {arXiv:1108.0602 [astro-ph.HE]} \BibitemShut
  {NoStop}%
\bibitem [{\citenamefont {Denicol}\ \emph {et~al.}(2019)\citenamefont
  {Denicol}, \citenamefont {Moln\'ar}, \citenamefont {Niemi},\ and\
  \citenamefont {Rischke}}]{Denicol:2019iyh}%
  \BibitemOpen
  \bibfield  {author} {\bibinfo {author} {\bibfnamefont {G.~S.}\ \bibnamefont
  {Denicol}}, \bibinfo {author} {\bibfnamefont {E.}~\bibnamefont {Moln\'ar}},
  \bibinfo {author} {\bibfnamefont {H.}~\bibnamefont {Niemi}}, \ and\ \bibinfo
  {author} {\bibfnamefont {D.~H.}\ \bibnamefont {Rischke}},\ }\href {\doibase
  10.1103/PhysRevD.99.056017} {\bibfield  {journal} {\bibinfo  {journal} {Phys.
  Rev. D}\ }\textbf {\bibinfo {volume} {99}},\ \bibinfo {pages} {056017}
  (\bibinfo {year} {2019})},\ \Eprint {http://arxiv.org/abs/1902.01699}
  {arXiv:1902.01699 [nucl-th]} \BibitemShut {NoStop}%
\bibitem [{\citenamefont {Kurian}\ and\ \citenamefont
  {Chandra}(2019)}]{Kurian:2019fty}%
  \BibitemOpen
  \bibfield  {author} {\bibinfo {author} {\bibfnamefont {M.}~\bibnamefont
  {Kurian}}\ and\ \bibinfo {author} {\bibfnamefont {V.}~\bibnamefont
  {Chandra}},\ }\href {\doibase 10.1103/PhysRevD.99.116018} {\bibfield
  {journal} {\bibinfo  {journal} {Phys. Rev. D}\ }\textbf {\bibinfo {volume}
  {99}},\ \bibinfo {pages} {116018} (\bibinfo {year} {2019})},\ \Eprint
  {http://arxiv.org/abs/1902.09200} {arXiv:1902.09200 [nucl-th]} \BibitemShut
  {NoStop}%
\bibitem [{\citenamefont {Hosoya}\ and\ \citenamefont
  {Kajantie}(1985)}]{Hosoya:1983xm}%
  \BibitemOpen
  \bibfield  {author} {\bibinfo {author} {\bibfnamefont {A.}~\bibnamefont
  {Hosoya}}\ and\ \bibinfo {author} {\bibfnamefont {K.}~\bibnamefont
  {Kajantie}},\ }\href {\doibase 10.1016/0550-3213(85)90499-7} {\bibfield
  {journal} {\bibinfo  {journal} {Nucl. Phys. B}\ }\textbf {\bibinfo {volume}
  {250}},\ \bibinfo {pages} {666} (\bibinfo {year} {1985})}\BibitemShut
  {NoStop}%
\bibitem [{\citenamefont {Ayala}\ \emph {et~al.}(2018)\citenamefont {Ayala},
  \citenamefont {Dominguez}, \citenamefont {Hernandez-Ortiz}, \citenamefont
  {Hernandez}, \citenamefont {Loewe}, \citenamefont {Manreza~Paret},\ and\
  \citenamefont {Zamora}}]{Ayala:2018wux}%
  \BibitemOpen
  \bibfield  {author} {\bibinfo {author} {\bibfnamefont {A.}~\bibnamefont
  {Ayala}}, \bibinfo {author} {\bibfnamefont {C.}~\bibnamefont {Dominguez}},
  \bibinfo {author} {\bibfnamefont {S.}~\bibnamefont {Hernandez-Ortiz}},
  \bibinfo {author} {\bibfnamefont {L.}~\bibnamefont {Hernandez}}, \bibinfo
  {author} {\bibfnamefont {M.}~\bibnamefont {Loewe}}, \bibinfo {author}
  {\bibfnamefont {D.}~\bibnamefont {Manreza~Paret}}, \ and\ \bibinfo {author}
  {\bibfnamefont {R.}~\bibnamefont {Zamora}},\ }\href {\doibase
  10.1103/PhysRevD.98.031501} {\bibfield  {journal} {\bibinfo  {journal} {Phys.
  Rev. D}\ }\textbf {\bibinfo {volume} {98}},\ \bibinfo {pages} {031501}
  (\bibinfo {year} {2018})},\ \Eprint {http://arxiv.org/abs/1805.08198}
  {arXiv:1805.08198 [hep-ph]} \BibitemShut {NoStop}%
\bibitem [{\citenamefont {Bandyopadhyay}\ \emph {et~al.}(2019)\citenamefont
  {Bandyopadhyay}, \citenamefont {Karmakar}, \citenamefont {Haque},\ and\
  \citenamefont {Mustafa}}]{Bandyopadhyay:2017cle}%
  \BibitemOpen
  \bibfield  {author} {\bibinfo {author} {\bibfnamefont {A.}~\bibnamefont
  {Bandyopadhyay}}, \bibinfo {author} {\bibfnamefont {B.}~\bibnamefont
  {Karmakar}}, \bibinfo {author} {\bibfnamefont {N.}~\bibnamefont {Haque}}, \
  and\ \bibinfo {author} {\bibfnamefont {M.~G.}\ \bibnamefont {Mustafa}},\
  }\href {\doibase 10.1103/PhysRevD.100.034031} {\bibfield  {journal} {\bibinfo
   {journal} {Phys. Rev. D}\ }\textbf {\bibinfo {volume} {100}},\ \bibinfo
  {pages} {034031} (\bibinfo {year} {2019})},\ \Eprint
  {http://arxiv.org/abs/1702.02875} {arXiv:1702.02875 [hep-ph]} \BibitemShut
  {NoStop}%
\bibitem [{\citenamefont {Hattori}\ \emph
  {et~al.}(2017{\natexlab{b}})\citenamefont {Hattori}, \citenamefont {Huang},
  \citenamefont {Rischke},\ and\ \citenamefont {Satow}}]{Hattori:2017qih}%
  \BibitemOpen
  \bibfield  {author} {\bibinfo {author} {\bibfnamefont {K.}~\bibnamefont
  {Hattori}}, \bibinfo {author} {\bibfnamefont {X.-G.}\ \bibnamefont {Huang}},
  \bibinfo {author} {\bibfnamefont {D.~H.}\ \bibnamefont {Rischke}}, \ and\
  \bibinfo {author} {\bibfnamefont {D.}~\bibnamefont {Satow}},\ }\href
  {\doibase 10.1103/PhysRevD.96.094009} {\bibfield  {journal} {\bibinfo
  {journal} {Phys. Rev. D}\ }\textbf {\bibinfo {volume} {96}},\ \bibinfo
  {pages} {094009} (\bibinfo {year} {2017}{\natexlab{b}})},\ \Eprint
  {http://arxiv.org/abs/1708.00515} {arXiv:1708.00515 [hep-ph]} \BibitemShut
  {NoStop}%
\bibitem [{\citenamefont {Strickland}(2014)}]{Strickland:2014pga}%
  \BibitemOpen
  \bibfield  {author} {\bibinfo {author} {\bibfnamefont {M.}~\bibnamefont
  {Strickland}},\ }\href {\doibase 10.5506/APhysPolB.45.2355} {\bibfield
  {journal} {\bibinfo  {journal} {Acta Phys. Polon. B}\ }\textbf {\bibinfo
  {volume} {45}},\ \bibinfo {pages} {2355} (\bibinfo {year} {2014})},\ \Eprint
  {http://arxiv.org/abs/1410.5786} {arXiv:1410.5786 [nucl-th]} \BibitemShut
  {NoStop}%
\bibitem [{\citenamefont {Schenke}\ \emph {et~al.}(2006)\citenamefont
  {Schenke}, \citenamefont {Strickland}, \citenamefont {Greiner},\ and\
  \citenamefont {Thoma}}]{Schenke:2006xu}%
  \BibitemOpen
  \bibfield  {author} {\bibinfo {author} {\bibfnamefont {B.}~\bibnamefont
  {Schenke}}, \bibinfo {author} {\bibfnamefont {M.}~\bibnamefont {Strickland}},
  \bibinfo {author} {\bibfnamefont {C.}~\bibnamefont {Greiner}}, \ and\
  \bibinfo {author} {\bibfnamefont {M.~H.}\ \bibnamefont {Thoma}},\ }\href
  {\doibase 10.1103/PhysRevD.73.125004} {\bibfield  {journal} {\bibinfo
  {journal} {Phys. Rev. D}\ }\textbf {\bibinfo {volume} {73}},\ \bibinfo
  {pages} {125004} (\bibinfo {year} {2006})},\ \Eprint
  {http://arxiv.org/abs/hep-ph/0603029} {arXiv:hep-ph/0603029} \BibitemShut
  {NoStop}%
\bibitem [{\citenamefont {Romatschke}\ and\ \citenamefont
  {Strickland}(2003)}]{Romatschke:2003ms}%
  \BibitemOpen
  \bibfield  {author} {\bibinfo {author} {\bibfnamefont {P.}~\bibnamefont
  {Romatschke}}\ and\ \bibinfo {author} {\bibfnamefont {M.}~\bibnamefont
  {Strickland}},\ }\href {\doibase 10.1103/PhysRevD.68.036004} {\bibfield
  {journal} {\bibinfo  {journal} {Phys. Rev. D}\ }\textbf {\bibinfo {volume}
  {68}},\ \bibinfo {pages} {036004} (\bibinfo {year} {2003})},\ \Eprint
  {http://arxiv.org/abs/hep-ph/0304092} {arXiv:hep-ph/0304092} \BibitemShut
  {NoStop}%
\bibitem [{\citenamefont {Schenke}\ and\ \citenamefont
  {Strickland}(2007)}]{Schenke:2006yp}%
  \BibitemOpen
  \bibfield  {author} {\bibinfo {author} {\bibfnamefont {B.}~\bibnamefont
  {Schenke}}\ and\ \bibinfo {author} {\bibfnamefont {M.}~\bibnamefont
  {Strickland}},\ }\href {\doibase 10.1103/PhysRevD.76.025023} {\bibfield
  {journal} {\bibinfo  {journal} {Phys. Rev. D}\ }\textbf {\bibinfo {volume}
  {76}},\ \bibinfo {pages} {025023} (\bibinfo {year} {2007})},\ \Eprint
  {http://arxiv.org/abs/hep-ph/0611332} {arXiv:hep-ph/0611332} \BibitemShut
  {NoStop}%
\bibitem [{\citenamefont {Srivastava}\ \emph {et~al.}(2015)\citenamefont
  {Srivastava}, \citenamefont {Thakur},\ and\ \citenamefont
  {Patra}}]{Srivastava:2015via}%
  \BibitemOpen
  \bibfield  {author} {\bibinfo {author} {\bibfnamefont {P.~K.}\ \bibnamefont
  {Srivastava}}, \bibinfo {author} {\bibfnamefont {L.}~\bibnamefont {Thakur}},
  \ and\ \bibinfo {author} {\bibfnamefont {B.~K.}\ \bibnamefont {Patra}},\
  }\href {\doibase 10.1103/PhysRevC.91.044903} {\bibfield  {journal} {\bibinfo
  {journal} {Phys. Rev. C}\ }\textbf {\bibinfo {volume} {91}},\ \bibinfo
  {pages} {044903} (\bibinfo {year} {2015})},\ \Eprint
  {http://arxiv.org/abs/1501.03576} {arXiv:1501.03576 [hep-ph]} \BibitemShut
  {NoStop}%
\bibitem [{\citenamefont {Kumar}\ \emph {et~al.}(2018)\citenamefont {Kumar},
  \citenamefont {Jamal}, \citenamefont {Chandra},\ and\ \citenamefont
  {Bhatt}}]{Kumar:2017bja}%
  \BibitemOpen
  \bibfield  {author} {\bibinfo {author} {\bibfnamefont {A.}~\bibnamefont
  {Kumar}}, \bibinfo {author} {\bibfnamefont {M.~Y.}\ \bibnamefont {Jamal}},
  \bibinfo {author} {\bibfnamefont {V.}~\bibnamefont {Chandra}}, \ and\
  \bibinfo {author} {\bibfnamefont {J.~R.}\ \bibnamefont {Bhatt}},\ }\href
  {\doibase 10.1103/PhysRevD.97.034007} {\bibfield  {journal} {\bibinfo
  {journal} {Phys. Rev. D}\ }\textbf {\bibinfo {volume} {97}},\ \bibinfo
  {pages} {034007} (\bibinfo {year} {2018})},\ \Eprint
  {http://arxiv.org/abs/1709.01032} {arXiv:1709.01032 [nucl-th]} \BibitemShut
  {NoStop}%
\bibitem [{\citenamefont {Khan}\ and\ \citenamefont
  {Patra}(2020)}]{Khan:2020rdw}%
  \BibitemOpen
  \bibfield  {author} {\bibinfo {author} {\bibfnamefont {S.~A.}\ \bibnamefont
  {Khan}}\ and\ \bibinfo {author} {\bibfnamefont {B.~K.}\ \bibnamefont
  {Patra}},\ }\href@noop {} {\  (\bibinfo {year} {2020})},\ \Eprint
  {http://arxiv.org/abs/2011.02682} {arXiv:2011.02682 [hep-ph]} \BibitemShut
  {NoStop}%
\bibitem [{\citenamefont {Jiang}\ \emph {et~al.}(2016)\citenamefont {Jiang},
  \citenamefont {Hou},\ and\ \citenamefont {Li}}]{Jiang:2016dkf}%
  \BibitemOpen
  \bibfield  {author} {\bibinfo {author} {\bibfnamefont {B.-f.}\ \bibnamefont
  {Jiang}}, \bibinfo {author} {\bibfnamefont {D.-f.}\ \bibnamefont {Hou}}, \
  and\ \bibinfo {author} {\bibfnamefont {J.-r.}\ \bibnamefont {Li}},\ }\href
  {\doibase 10.1103/PhysRevD.94.074026} {\bibfield  {journal} {\bibinfo
  {journal} {Phys. Rev. D}\ }\textbf {\bibinfo {volume} {94}},\ \bibinfo
  {pages} {074026} (\bibinfo {year} {2016})}\BibitemShut {NoStop}%
\bibitem [{\citenamefont {Carrington}\ \emph {et~al.}(2004)\citenamefont
  {Carrington}, \citenamefont {Fugleberg}, \citenamefont {Pickering},\ and\
  \citenamefont {Thoma}}]{Carrington:2003je}%
  \BibitemOpen
  \bibfield  {author} {\bibinfo {author} {\bibfnamefont {M.}~\bibnamefont
  {Carrington}}, \bibinfo {author} {\bibfnamefont {T.}~\bibnamefont
  {Fugleberg}}, \bibinfo {author} {\bibfnamefont {D.}~\bibnamefont
  {Pickering}}, \ and\ \bibinfo {author} {\bibfnamefont {M.}~\bibnamefont
  {Thoma}},\ }\href {\doibase 10.1139/p04-035} {\bibfield  {journal} {\bibinfo
  {journal} {Can. J. Phys.}\ }\textbf {\bibinfo {volume} {82}},\ \bibinfo
  {pages} {671} (\bibinfo {year} {2004})},\ \Eprint
  {http://arxiv.org/abs/hep-ph/0312103} {arXiv:hep-ph/0312103} \BibitemShut
  {NoStop}%
\bibitem [{\citenamefont {Aarts}\ \emph {et~al.}(2007)\citenamefont {Aarts},
  \citenamefont {Allton}, \citenamefont {Foley}, \citenamefont {Hands},\ and\
  \citenamefont {Kim}}]{Aarts:2007wj}%
  \BibitemOpen
  \bibfield  {author} {\bibinfo {author} {\bibfnamefont {G.}~\bibnamefont
  {Aarts}}, \bibinfo {author} {\bibfnamefont {C.}~\bibnamefont {Allton}},
  \bibinfo {author} {\bibfnamefont {J.}~\bibnamefont {Foley}}, \bibinfo
  {author} {\bibfnamefont {S.}~\bibnamefont {Hands}}, \ and\ \bibinfo {author}
  {\bibfnamefont {S.}~\bibnamefont {Kim}},\ }\href {\doibase
  10.1103/PhysRevLett.99.022002} {\bibfield  {journal} {\bibinfo  {journal}
  {Phys. Rev. Lett.}\ }\textbf {\bibinfo {volume} {99}},\ \bibinfo {pages}
  {022002} (\bibinfo {year} {2007})},\ \Eprint
  {http://arxiv.org/abs/hep-lat/0703008} {arXiv:hep-lat/0703008} \BibitemShut
  {NoStop}%
\bibitem [{\citenamefont {Satow}(2014)}]{Satow:2014lia}%
  \BibitemOpen
  \bibfield  {author} {\bibinfo {author} {\bibfnamefont {D.}~\bibnamefont
  {Satow}},\ }\href {\doibase 10.1103/PhysRevD.90.034018} {\bibfield  {journal}
  {\bibinfo  {journal} {Phys. Rev. D}\ }\textbf {\bibinfo {volume} {90}},\
  \bibinfo {pages} {034018} (\bibinfo {year} {2014})},\ \Eprint
  {http://arxiv.org/abs/1406.7032} {arXiv:1406.7032 [hep-ph]} \BibitemShut
  {NoStop}%
\bibitem [{\citenamefont {Bjorken}(1983)}]{Bjorken:1982qr}%
  \BibitemOpen
  \bibfield  {author} {\bibinfo {author} {\bibfnamefont {J.}~\bibnamefont
  {Bjorken}},\ }\href {\doibase 10.1103/PhysRevD.27.140} {\bibfield  {journal}
  {\bibinfo  {journal} {Phys. Rev. D}\ }\textbf {\bibinfo {volume} {27}},\
  \bibinfo {pages} {140} (\bibinfo {year} {1983})}\BibitemShut {NoStop}%
\end{thebibliography}%

\end{document}